\documentclass[10pt]{article} 
\usepackage[preprint]{tmlr}
\usepackage{hyperref}
\usepackage{url}

\usepackage{graphicx}
\usepackage{siunitx}  
\usepackage{amsmath, amssymb}

\def\month{11}  
\def\year{2026} 
\def\openreview{\url{https://openreview.net/forum?id=zKLMev1mnQ}}

\begin{document}

\title{Spectral and Conformal: Neural Network Methods for Calibrated Trajectory Prediction and Control-Point Intervals}
\date{October 2026}
\author{\name Jasmine Ratchford \email jasmine.star.ma@gmail.com \addr  CMU Software Engineering Institute (former affiliation) \AND
\name Raffaele Romagnoli \email rromagnoli@duq.edu 
\addr Duquesne University
}


\maketitle
\begin{abstract}

Autonomous systems that navigate their environment need methods to quickly assess the feasibility and safety of desired trajectories. Traditional solutions that rely on numerical integration can be computationally heavy, while fast data-driven surrogates lack the formal guarantees that numerical approaches provide. This work demonstrates a framework that uses (1) a neural network with a spectral backbone to learn trajectories and quantile bands with built-in constraints (initial conditions and stability); (2) a modified conformal quantile regression to provide calibrated, joint trajectory-level coverage; and (3) interval arithmetic to extend noise-free
coverage guarantees to control points. 
After refactoring the network for inference, the trajectory inference time for a quadrotor under a closed-loop linear quadratic regulator is 3.45~ms on a Raspberry Pi Zero 2 W. The total pipeline time, including trajectory inference time, is 9.81~ms, or approximately 100~Hz, consistent with typical control rates. 

\end{abstract}
\section{Introduction}

Autonomous systems that navigate their environment need methods to quickly assess the feasibility and safety of trajectories. This challenge arises in a range of applications where safety constraints must be satisfied despite nonlinear dynamics, disturbances, and modeling uncertainty (for example, autonomous vehicles and aerial robots). Techniques that address this challenge, such as reachability analysis \citep{bansal2017hamilton}, 
reference governors \citep{garone2017reference}, 
and various predictive control techniques \citep{rawlings2017model}, rely on accurate models of the system dynamics and are often dependent on numerical integration that is computationally demanding, especially for highly nonlinear systems \citep{wabersich2023data}.

Replacing numerical integration with a data-driven surrogate model is a common solution \citep{giorgiani2020quadcopter,safe_fast_mpc_nn_2020}. 
However, because surrogates lack the guarantees of full numerical integration and can carry unquantified, unassessed uncertainties, their use in safety-critical applications remains challenging. Furthermore, detailed trajectory representations are often not in a form suitable for safety verification or online decision making. 


Here, we develop a framework that combines trajectory learning with spline-based representations to generate uncertainty-quantified trajectory predictions with noise-free coverage guarantees in a compact representation. The spline representation is useful because B-spline trajectories are convex combinations of their control points. Therefore, safety constraints can be directly evaluated using control-point sets. With a surrogate model, such estimation can occur without repeated numerical integration of the underlying dynamics. 

We apply conformal quantile regression (CQR) \citep{romano2019conformalized} to provide per-dimension, per-time step, and joint (total trajectory) prediction coverage and show that splines can inherit the uncertainty ranges of predicted trajectories. 
Our benchmarking studies on a Raspberry Pi Zero 2 W show that real-time safety assessment is feasible with this method. Overall, this work demonstrates a neural-network-and-CQR-based pipeline to acquire adaptive-knot B-spline control intervals from initial conditions for an autonomous ordinary differential equation (ODE), 
without numerical integration. The primary contributions of this work are: 
\begin{itemize}
\item A method for producing calibrated, verification-ready uncertainty: conformal quantile regression (CQR) yields \emph{whole-trajectory} (joint) coverage guarantees under noise-free conditions, 
and the resulting quantile bands map to B-spline control-point ranges via interval arithmetic. \item A spectral decomposition layer (where a shared set of learned eigenvalues governs trajectory evolution) combined with a triplet of decoders that together provide smooth quantile estimates.  
\item Inference benchmarks on constrained hardware showing latency within a 100~Hz control-rate budget.

\end{itemize}

\section{Preliminaries}
\paragraph{Notation.}
Throughout, $f(t)$ denotes the true trajectory of the system: the solution of the underlying ODE from an initial condition $x_0 \in \mathbb{R}^{12}$ (12 is the number of dimensions of the LQR equations), sampled at $m$ rollout points. The network maps $x_0$ to a predicted trajectory $x(t)$, together with lower and upper quantile estimates $x_{\text{lo}}(t)$ and $x_{\text{hi}}(t)$. By construction, $x(t=0) = x_0$. Note that $t$ is actually a sampled set $t\in\mathcal{T}= \{0,...,t_m\}$ of the $m$ rollout points. 

A set of B-spline basis functions $B \in \mathbb{R}^{m \times n_{cp}}$ is constructed for a given set of knot locations $\{t_j\}$, where $n_{cp}$ denotes the number of control points: for a clamped knot vector with $n_{\text{int}}$ interior knots and degree $k$, $n_{cp} = n_{\text{int}} + k + 1$ (for cubic splines, $n_{\text{int}} + 4$). Throughout, $i \in \{1, \dots, n_{cp}\}$ indexes basis functions and control points, and $j \in \{1, \dots, n_{\text{int}}\}$ indexes interior knots. A B-spline of degree $k$ with basis functions $B^k_i(t)$, determined by a knot vector, represents a trajectory through control points: $s(t) = B(t)\,c$ denotes the spline reconstruction of the true trajectory, with $c = B^\dagger f$, while $\hat{s}(t) = B(t)\,\hat{c}$, with $\hat{c} = B^\dagger x$, denotes the spline reconstruction of the \emph{predicted} trajectory. The pipeline outputs $\hat{s}(t)$; $s(t)$ serves as its idealized counterpart.

\paragraph{Treatment and Framing.} 
We treat the trajectory $f(t)$ as any continuous function approximated by a spline. Given a knot vector, the spline approximation reduces to a fast approximation of the trajectory itself, $x(t)$. Therefore, rather than predicting the sampled $f(t)$ directly, we predict a spline approximation $\hat{s}(t)$. 
Thus the primary effort is estimating the full trajectory, with control point calculation a fast, post-prediction step. 
Convergence to an equilibrium does not by itself guarantee safety along the transient trajectory. The purpose of the proposed framework is to complement the stability of the controlled system with a probabilistic assessment of transient constraint satisfaction. Under the stated conformal assumptions, containment of the calibrated trajectory envelope within the safe set provides a probabilistic safety assessment of the sampled trajectory.
We place knots adaptively, using the relationship between knot density, local curvature, and maximum error \citep{de1978practical,ongoing} to avoid unnecessarily dense knot vectors and the pseudoinverse burden they incur. This step is optional.

Notably, a Koopman operator framing recasts a nonlinear dynamical system in terms of an infinite-dimensional lifted space of observable functions that evolve \emph{linearly} under the dynamics; finite-dimensional approximations of this lifted evolution are commonly learned via autoencoder architectures \citep{lusch2018deep,pan2020physics, otto2021koopman,brunton2022modern}. Our spectral layer can be viewed similarly, but without an autoencoder architecture. The methods discussed are designed for autonomous ordinary differential equations (ODEs) whose trajectories are uniquely determined from an initial condition. Ideally, the system is stable or the time horizon is short enough that short-time linearization is valid and the trajectory can be approximated by a spline with reasonable knot point density and low degree. For this, we demonstrate using a nonlinear quadrotor system controlled via linear quadratic regulator (LQR), as described in Appendix~\ref{app:dynamics}. 

\section{Related Work}
Learning 
Koopman or Lie operators through neural networks is well studied \citep{lusch2018deep,brunton2016koopman,brunton2022modern,otto2019linearly,otto2021koopman}. Specifically, the architecture employed in this paper is similar to 
\cite{pan2020physics} and \cite{jamil2026lie}. 
We differ from \cite{jamil2026lie} in that our architecture does not include an inverse fast Fourier transform (IFFT) and lacks the spatial distribution inherent in their partial differential equation (PDE) setting. 
We enforce stability by directly capping eigenvalues rather than through negative-semi-definite parameterization used in \cite{pan2020physics} or skew-symmetric decomposition such as in \cite{jamil2026lie}. Furthermore, we quantify uncertainty via conformal calibration rather than Bayesian variational inference. 
The model in \cite{otto2019linearly} is a discrete-time system whereas ours is treated as continuous.

Pinball loss is naturally suited to conformal calibration. Exchangeability
is easily satisfied because training, calibration, and test data are drawn as independent and identically distributed samples from the same input distribution. Applied per-dimension and per-time step, conformalized quantile regression (CQR) \citep{romano2019conformalized} yields our cell- and step-level guarantees directly. At the trajectory level, the useful guarantee is joint: coverage holding simultaneously across all time steps and dimensions. Joint coverage can be obtained in several ways \citep[see][]{stankeviciute2021conformal,ICLR2024_sun_copula}. We instead follow the supremum-based construction of \citet{diquigiovanni2022conformal}. In order to express the predicted quantile range using splines, we 
map the bounds through the pseudoinverse. This is distinct from common forward treatments such as \cite{van2015b} and \cite{rousseau2019minimum} that use pointwise constraints derived from system dynamics to build constraints on the control-point values. Structurally, our approach to calculating one calibration scalar, $Q$, from multidimensional data is similar to \citet{diquigiovanni2022conformal} and \citet{dietterich2022conformal}, both of which leverage correlation between dimensions. However, the inherent smoothness of autonomous ODEs and our neural network approach differ sharply from the discrete nature of the techniques and systems explored in \citet{dietterich2022conformal}.

\section{Methodology}
We describe our methodology here, which has two components.
\begin{itemize}
    \item[1.]Trajectory prediction is performed using a neural network (75K parameters) that takes as input the d-dimensional initial condition $x_0^d$. This has three sub-components: a fully connected neural network that acts as a nonlinear encoder ("lift" operation); a spectral layer; and a triplet of decoders that estimate the mean and quantile range. This is shown in Figure~\ref{fig:diagram}. 
    \item[2.] A post-network CQR analysis is combined with B-splines to enable a compact representation of results with quantified (noise-free) estimation uncertainty.
\end{itemize}
The full code used in this project is available at \url{https://github.com/cmu-sei/bspline_net}.

\subsection{Trajectory Prediction}

\begin{figure}[ht]
    \centering
    \includegraphics[width=0.9\linewidth]{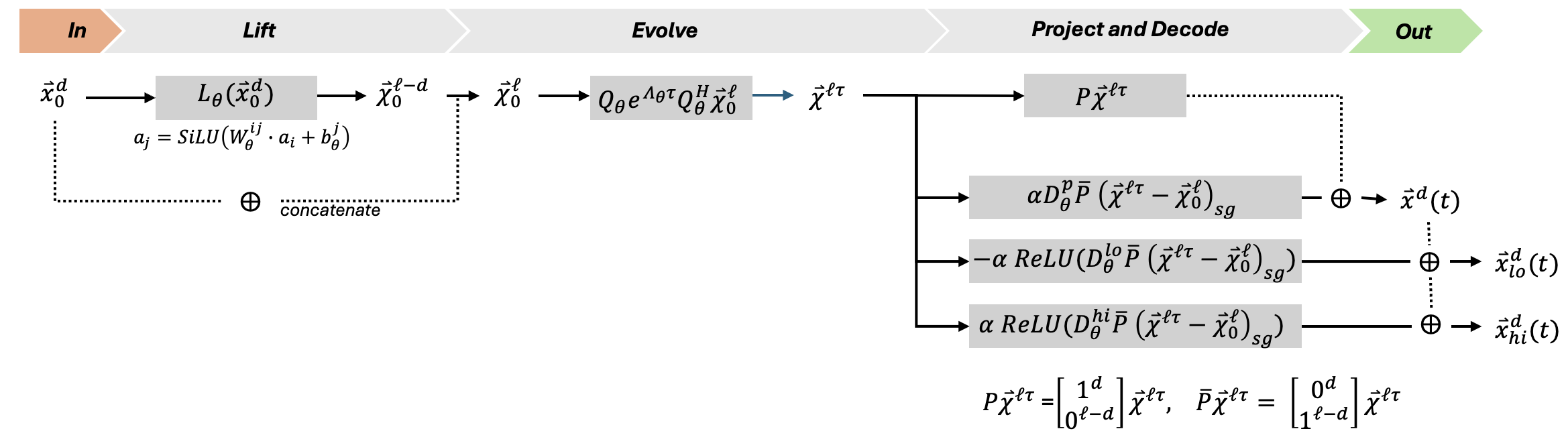}
    \caption{A schematic of the trajectory network showing the lift, evolve, and project and decode portions of the network in sequence. The initial condition $\vec{x}_0^{\,d}$ is concatenated with the output of a SiLU encoder $L_\theta$ to form the lifted state $\vec{\chi}_0^{\,\ell}$, which the spectral layer $Q_\theta e^{\Lambda_\theta \tau} Q_\theta^H$ evolves linearly to any query time $\tau$. The projection $P\vec{\chi}^{\,\ell\tau}$ plus a mean-correction head yields $\vec{x}^{\,d}(t)$; two ReLU-gated quantile heads, added around the mean, yield $\vec{x}_{\text{lo}}^{\,d}(t)$ and $\vec{x}_{\text{hi}}^{\,d}(t)$. All heads act on the stop-gradient (sg) residual $\bar{P}(\vec{\chi}^{\,\ell\tau} - \vec{\chi}_0^{\,\ell})$, with gate $\alpha$ enabled after warmup; at $\tau{=}0$ the output equals $x_0$
by construction.}
    \label{fig:diagram}
\end{figure}

The model maps an initial condition $x_0 \in \mathbb{R}^{d}$ to a predicted trajectory through three stages: a lift into a higher-dimensional observable space, a stable continuous-time spectral evolution, and an anchored quantile decoder. Hereafter, $x(t)\equiv x^d_{\text{mean}}(t)$ denotes the point prediction.
\paragraph{Lift.} The initial condition is lifted by a fully connected neural network  $L_\theta$ and
concatenated with the initial condition state itself,
\begin{equation}
  {\chi}_0^{\ell} = x_0^{d} \oplus L_\theta(x_0^{d}),
  \qquad
  a_j = \mathrm{SiLU} \!\big(W_\theta^{ij} a_i + b_\theta^{j}\big),
\end{equation}
where $\oplus$ denotes concatenation, $d$ is the state dimension, and $\ell$ the lifted dimension. The concatenated initial condition ensures that no information is lost in the lift operation. 
The $\mathrm{SiLU}$ activation function was selected for internal layers because of its smooth, non-saturating gradient. 
\paragraph{Evolve.} The lifted state is propagated through the set of times, $\tau$, in closed form through the stable spectral layer,
\begin{equation}
  {\chi}^{\ell\tau} = Q_\theta\, e^{\Lambda_\theta \tau}\, Q_\theta^{H}\, {\chi}_0^{\ell},
\end{equation}
with $Q_\theta = \exp(S),\ S = A_\theta-A_\theta^{H}$ and $\mathrm{Re}\,\Lambda_\theta \le -\epsilon_{\min} < 0$ enforcing stability. $\tau$ is the set of sampled times at which ${\chi}^{\ell\tau} $ is evaluated and is not learned. $ Q_\theta\,$ and $\Lambda_\theta$ are learned parameters ($ Q_\theta\,$ is learned via $A_\theta$) with $^H$ representing the Hermitian conjugate. ${\chi}^{\ell\tau} $ is in the lifted space. 

\paragraph{Project and Decode.}

By construction, ${\chi}_0^{\ell}$ contains the initial condition as a subspace (via concatenation) and selecting those coordinates is exactly correct at $\tau=0$. We define the projection operators, $P$ and $\bar{P}$, as the operations that project the output of the spectral layer onto the physical state subspace and its complementary subspace, respectively. 
Because $P\chi^{\ell\tau}$ is linear in the lifted state, it captures the easily learned linear evolution, which forms the backbone for nonlinear refinement. Decoders act only on the \emph{non-physical} subspace of the evolved lifted state, detached from the computational graph to prevent distortion of the linear backbone. Then, the input to the decoders is:
\begin{equation}
  \tilde{h}(\tau) = \mathrm{sg}\!\big[\, \bar{P}\,(\vec\chi^{\ell\tau} -\vec\chi^{\ell}_0)\,\big],
\end{equation}
where $\mathrm{sg}[\cdot]$ denotes stop-gradient. Each decoder is a single learnable matrix without a bias term. Because $\tilde h(0)=0$, there is no correction at $\tau=0$. The mean and quantile band low-and-high predictions are then:
\begin{align}
  x^d_{\text{mean}}(\tau) &= P\,{\chi}^{\ell\tau}
      + \alpha D_\theta^{\text{mean}}\tilde{h}(\tau) , \label{eq:decoder12} \\
  x^d_{\text{lo}}(\tau) &= x^d_{\text{mean}}(\tau)
      - \alpha\,\mathrm{ReLU}\!\big(D_\theta^{\text{lo}}\tilde{h}(\tau) \big), \label{eq:decoder3}\\
  x^d_{\text{hi}}(\tau) &= x^d_{\text{mean}}(\tau)
      + \alpha\,\mathrm{ReLU}\!\big(D_\theta^{\text{hi}}\tilde{h}(\tau) \big).\label{eq:decoder4}
\end{align}
The $\mathrm{ReLU}$ final activation function ensures non-negative offsets from the mean and maintains the order of the quantile range by construction.  
 For the mean prediction, this architecture propagates lifted-state perturbations to state-space
prediction errors through a single linear operator,
\begin{equation}
  M(\tau) = P\,Q_\theta\, e^{\Lambda_\theta \tau}\, Q_\theta^{H}
          + \alpha\,D^{\text{mean}}_\theta\bar{P}\!\left(Q_\theta\, e^{\Lambda_\theta \tau}\, Q_\theta^{H} - I\right).
\end{equation}
At $\tau = 0$, $Q_\theta e^{\Lambda_\theta \cdot 0} Q_\theta^{H} = I$, so $M(0) = P$: the prediction error and band width are zero at the initial condition by construction. For $\tau > 0$, the evolved component $Q_\theta e^{\Lambda_\theta \tau} Q_\theta^{H}$ decays at a rate set by $\mathrm{Re}\,\Lambda_\theta$. We enforce $\mathrm{Re}\,\Lambda_\theta \le -\epsilon_{\min}$ by parameterization to ensure stability.
A diagram of the network is shown in Figure~\ref{fig:diagram}.

\paragraph{Loss.}
The training objective combines a mean trajectory loss, two pinball (quantile) losses, and an optional integral-consistency penalty. For the sampled true trajectory $f$ and pinball loss $\rho_q(u) = \max(q u,\,(q-1)u)$ at quantile level $q$,

\begin{equation}
  \mathcal{L} =
  \underbrace{\mathcal{L}_{\mathrm{traj}}
            + \mathcal{L}_{\mathrm{traj}}^{\mathrm{norm}}}_{\text{mean fit}}
  + \underbrace{\rho_{q_{\mathrm{lo}}}\!\big(f - x^d_{\text{lo}}\big)
              + \rho_{q_{\mathrm{hi}}}\!\big(f - x^d_{\text{hi}}\big)}_{\text{quantile bands}}
  + \lambda_{\mathrm{int}}\,\alpha\,\mathcal{L}_{\mathrm{int}}\label{eq:loss}
\end{equation}
where the mean terms are the mean squared error and its per-channel standard-deviation--normalized counterpart,
\begin{equation}
  \mathcal{L}_{\mathrm{traj}}
    = \frac{1}{d|\tau|}\big\| x^d_{\text{mean}} - f \big\|_2^2,
  \qquad
  \mathcal{L}_{\mathrm{traj}}^{\mathrm{norm}}
    = \frac{1}{d|\tau|}\Big\| \big(x^d_{\text{mean}} - f\big) \oslash \sigma \Big\|_2^2,\label{eq:spec_loss}
\end{equation}
with $\sigma$ the per-channel standard deviation and $\oslash$ element-wise division, and $|\tau|$ the number of sampled times. The integral-consistency term enforces the kinematic relation that position increments match the trapezoidal integral of velocity (in physical units), and is active only when $\lambda_{\mathrm{int}} > 0$. Notably, because $x^d_{\text{lo}}$ and $x^d_{\text{hi}}$ include $\alpha$-dependent adjustments to the mean value, they are fixed to $x^d_{\text{mean}}$ until $\alpha$ turns on.

\subsection{Deriving Quantiles}
We use \emph{Conformal Quantile Regression} (CQR) as proposed by \cite{romano2019conformalized} to evaluate and scale the quantiles predicted by our model to guarantee coverage. The general CQR procedure is summarized by three steps: 
\begin{itemize}
    \item [1] Generate data split into a calibration and hold-out dataset and predict the quantile bounds (($L_{low}, L_{high}$), in our case using ($x^d_\text{lo}, x^d_\text{hi}$)) using the calibration set. 
    \item [2] Compute the score $\eta = \max(L_{low} - f, f - L_{high})$, which quantifies whether each sample is within the quantile band  ($\eta<0$) or not ($\eta>0$). 
    \item [3] Create a new calibrated band ($L_{low}-Q_\eta, L_{high}+Q_\eta$)  where $Q_\eta$ is the $(1-\delta)(1+1/n)$ quantile of $\eta$ ($n$ is the number of calibration samples and in practice $1/n\ll 1$). \cite{romano2019conformalized} show this for symmetric and asymmetric quantiles, but we focus on symmetric quantiles in this paper.

\end{itemize}
Control problems necessitate total trajectory guarantees, which require \emph{joint coverage}.  Thus, we modify this procedure in two ways closely following the supremum-based construction by \citet{diquigiovanni2022conformal}. First, we generate a separate set of predictions to measure the per-dimension accuracy by calculating the absolute error distribution ($L_1$ distance from the \emph{point prediction}) for each dimension independently and reporting its 95\% quantile. We chose the 95\% quantile both to match the quantile of interest for the overall procedure and because we wanted the scaling to be sensitive to the tail of the distribution. The score in step (2) is computed on the trajectories scaled by $1/(\vec{R}\odot\vec{\omega})$ where $\vec{\omega}$ is the $L_1$ scaling. To recover the correct units, the conformal factor $Q_\eta$ in step (3) is multiplied by $\vec{R}\odot\vec{\omega}$. A primary motivation for this is that our model predicts in a normalized coordinate system where each dimension is scaled by the range of initial conditions such that the 12-D ball of initial conditions has unit radius. Physically, this is arbitrary. Without a rescaling prior to the CQR procedure, the resulting scale is influenced by the dimension that has the largest variance under the arbitrary scaling. 

Second, we modify the score in step (2) to be the \emph{maximum} value for any dimension or time step.  Because the score now equals the per-trajectory worst-case cell, the conformal guarantee in step (3) applies jointly across the trajectory. In our analysis, we used 100,000 trajectories total. 40,000 trajectories were used to obtain the $\omega$ scaling and another 40,000 trajectories to obtain $Q_\eta$. A final 20,000 trajectories were saved as a hold-out set for evaluation.  

\subsection{Extending quantiles to control point intervals} \label{sec:method_extend_quantiles}

Thus far we have demonstrated a coverage-guaranteed estimate of the trajectory. We return to the problem of providing a compact spline representation, where the point estimate is found simply via $c^d = B^\dagger \cdot x^d_\text{mean}$. It is not obvious how the quantile coverage can be incorporated into this representation: naively spline-fitting the upper and lower bands separately gives spline-encoded curves, but not the valid range for each control point of the mean spline. We suggest the following method, which finds the range of control-point values admissible by the set of functions within the quantile range.

 
The conformal calibration procedure described in the previous section produces, for each state component $d$, a prediction interval with joint coverage at least $1-\delta$; that is, the true trajectory $f^{d}$ satisfies
$x_{\text{lo}}^{d}(t) \le f^{d}(t) \le x_{\text{hi}}^{d}(t)$ for all
$t \in \mathcal{T}$ with probability at least $1-\delta$. Let
\begin{equation}
\mathcal F_{in}^{d} = \left\{g^{d} :
 x_{\text{lo}}^{d}(t) \le g^{d}(t) \le x_{\text{hi}}^{d}(t),
 \quad \forall t\in\mathcal T \right\}
\end{equation}
denote the set of admissible trajectories for component $d$. Given the knot vector $\{t_j\}$, let $B$ denote the corresponding B-spline collocation matrix evaluated at the sampling instants. For any $g^{d}\in\mathcal F_{in}^{d}$, the associated control points are $c^{d} = B^{\dagger}g^{d}$, where $B^{\dagger}$ denotes the Moore--Penrose pseudoinverse. 

Our goal is to determine an interval containing all admissible values of $c_i^{d}$ induced by the trajectory uncertainty band. We define $c_{i,\text{hi}}^{d} $ and $c_{i,\text{lo}}^{d}$ as  
\begin{align}
c_{i,\text{hi}}^{d} = \sup_{g^{d}\in\mathcal F_{in}^{d}} c_i^{d}(g^{d}),
\qquad
c_{i,\text{lo}}^{d} = \inf_{g^{d}\in\mathcal F_{in}^{d}} c_i^{d}(g^{d}).
\label{eq:cp_upper_lower}
\end{align}

Since $c_i^{d}$ is linear in $g^{d}$, the supremum and infimum over $\mathcal F_{in}^{d}$ decouple across sampling instants. The maximizing trajectory selects $x_\text{hi}^{d}(t)$ where $B_{it}^{\dagger}>0$ and $x_\text{lo}^{d}(t)$ where $B_{it}^{\dagger}<0$, and symmetrically for the minimizer. Writing $(B_{it}^{\dagger})^{\pm} = \max(\pm B_{it}^{\dagger}, 0)$, the extremal control-point values are

\begin{equation}
c_{i,\text{hi/lo}}^{d} = \sum_{t\in\mathcal T} (B_{it}^{\dagger})^{+}\, x_{\text{hi/lo}}^{d}(t) \;-\; (B_{it}^{\dagger})^{-}\, x_{\text{lo/hi}}^{d}(t).
\label{eq:cp_hi_lo}
\end{equation}


Every admissible trajectory $g^{d}\in\mathcal F_{in}^{d}$ therefore generates control points satisfying $c_{i,\text{lo}}^{d} \le c_i^{d} \le c_{i,\text{hi}}^{d}$. These bounds admit a compact matrix formulation. Define the midpoint (mid) and half-width (hw) for both the prediction interval and control points:

\begin{align}
x_{\text{mid}}^{d}(t) &= \frac{x_{\text{hi}}^{d}(t) + x_{\text{lo}}^{d}(t)}{2}, &
x_{\text{hw}}^{d}(t) &= \frac{x_{\text{hi}}^{d}(t) - x_{\text{lo}}^{d}(t)}{2}&
c_{i,\text{mid}}^{d} &= \frac{c_{i,\text{hi}}^{d} + c_{i,\text{lo}}^{d}}{2}, &
c_{i,\text{hw}}^{d} &= \frac{c_{i,\text{hi}}^{d} - c_{i,\text{lo}}^{d}}{2}.
\end{align}

Using $|B^\dagger_{it}|=(B^\dagger_{it})^++(B^\dagger_{it})^-$ and $B^\dagger_{it}=(B^\dagger_{it})^+-(B^\dagger_{it})^-$, adding and subtracting \eqref{eq:cp_hi_lo}'s two cases gives



\begin{equation}
c_{\text{mid}}^{d} = B^{\dagger}\, x_{\text{mid}}^{d}, \qquad c_{\text{hw}}^{d} = |B^{\dagger}|\, x_\text{hw}^{d}.
\label{eq:cp_matrix}
\end{equation}

Equation~\eqref{eq:cp_matrix} is the standard midpoint--radius form of interval arithmetic for linear maps \citep[see, e.g.,][]{moore2009introduction, neumaier1990interval}, applied here to the spline pseudoinverse. The interval midpoints are the least-squares projection of the band midpoint, exactly as in the point-estimate case, while the interval widths are linear in the band half-widths through the entrywise non-negative matrix $|B^{\dagger}|$. Zero trajectory uncertainty yields zero control-point uncertainty. The structure of $|B^{\dagger}|$ — in particular, which of its entries are large — determines which control points are most sensitive to trajectory uncertainty.


Because $\mathcal{F}_{in}$ contains the true trajectory with
probability at least $1-\delta$, the resulting control-point intervals
contain the control points of its discrete least-squares spline fit
with at least the same probability. Let
\[
    P_{\mathcal T}f(t)
    :=
    B(t)B^\dagger\boldsymbol f,
    \qquad
    \boldsymbol f=[f(t)]_{t\in\mathcal T},
\]
denote the discrete projector, and define its stability constant as
\[
    \Lambda_{\mathcal T}
    :=
    \sup_t\|B(t)B^\dagger\|_1.
\]
Provided that $B$ has full column rank, a condition that can be
established from the sampling points and knot vector through the
Schoenberg--Whitney conditions \citep{de1978practical},
$P_{\mathcal T}$ reproduces the selected spline space $\mathcal S$. A standard projector argument,
combined with the Jackson-type best-approximation bound
\citep{devore1998nonlinear,ongoing}, therefore gives
\begin{equation}\label{eq:deboor_bound}
\begin{aligned}
    \|f-P_{\mathcal T}f\|_\infty
    &\le
    (1+\Lambda_{\mathcal T})
    \inf_{v\in\mathcal S}\|f-v\|_\infty \\
    &\le
    (1+\Lambda_{\mathcal T})C_d
    \max_j h_j^2
    \|\ddot f\|_{\infty,[t_j,t_{j+1}]}
    =:\epsilon_{\mathrm{knot}},
\end{aligned}
\end{equation}
where $h_j$ is the local knot spacing, $C_d$ depends only on the
spline degree, and
$\Lambda_{\mathcal T}\leq\|B^\dagger\|_\infty$ characterizes the
stability of the sampled least-squares projector. The control-point guarantee established below applies to the least-squares spline approximation of the true trajectory. Extending this guarantee to the true continuous trajectory requires tightening the constraints by the spline-approximation residual characterized in Equation~\eqref{eq:deboor_bound}.

Setting aside $\epsilon_\text{knot}$ momentarily: because B-spline curves lie within the convex hull of their control points, any convex constraint satisfied at every point of the control-point intervals is satisfied by the corresponding spline functions. The control point intervals contain the \emph{spline fit of} the true trajectory, and thus the spline fit satisfies the convex constraint with probability at least $1-\delta$. This is a conservative estimate for two reasons. First, the mapping from $\mathcal{F}_{in}^{d}$ to control-point intervals is not invertible: the resulting set of control-point intervals contains $B^{\dagger}\mathcal{F}_{in}^{d}$ as a strict subset and thus admits control point vectors that generate functions that lie outside the quantile bands. Second, the convex hull represents a loose outer containment that is dependent on the number of control points \citep{de1978practical}. 
In the future, we may be able to extend the guarantee to the true trajectory itself: a reference declared admissible is then safe with probability at least $1-\delta$.  In practice, obtaining a physics-driven $\epsilon_{knot}$ requires deriving a rigorous bound on trajectory curvature from the system's known dynamics and actuator limits (unknown in this study). 

Notably, imposing smoothness constraints on the admissible set is a promising direction to further tighten the control point intervals. Restricting $\mathcal{F}_{in}^{d}$ to functions with bounded increments, $|g^{d}(t_{a+1}) - g^{d}(t_a)| \le L^{d}\,\Delta t_a$, excludes non-physical extremizers, and the extremization is no longer decoupled over sampling instants. The rate bound $L^{d}$ need not be assumed: evaluating the system dynamics over the extreme values of the conformal band yields a certified bound on $|\dot{f}^{d}|$. The true trajectory remains in the restricted set whenever it lies in the band, and the $1-\delta$ guarantee transfers to the tighter intervals.

\section{Experiments}
\subsection{Baseline Trajectory Prediction}
For our experiments in trajectory prediction, we used the system as described in Appendix~\ref{app:dynamics}. For each initial condition, a trajectory is generated via numerical integration of the nonlinear dynamics equations. For subsequent spline approximations, we use cubic B-splines. The physical constants we use in the quadrotor solver are given in Table~\ref{tab:dynamics}.  Inputs are normalized by a radius $R$, which is set by the outer perimeter of the ball of initial conditions, but more generally is arbitrary. 

The baseline model is described in Table~\ref{tab:traj_network} in Appendix~\ref{app:training}. The learnable parameters are approximately evenly distributed between the Lift and Evolve steps with the three Decoder layers' parameters accounting for about 12\% of the total learnable parameters. Empirically, we found that the values used here balanced network size and performance. More details are available in Appendix~\ref{app:ablation}.


\paragraph{Data generation.} Data were generated using initial conditions randomly selected from within the ball of allowable initial conditions given in Table~\ref{tab:dynamics} (weighted toward initial conditions with larger radius). We numerically integrated the dynamics using a fourth-order Runge--Kutta (RK4) scheme, JIT-compiled with \texttt{Numba}.  

\paragraph{Training procedure.} The gate $\alpha$ follows a linear warm-up schedule, held at $0$ until epoch $e_{\mathrm{start}}= 400$, ramped linearly to $1$ by epoch $e_{\mathrm{end}}=500$, and fixed at $1$ thereafter, so the decoder correction is introduced only after the spectral backbone has settled. The spectral parameters use a learning-rate multiplier of $3$ relative to the rest of the network, and the learning rate begins at $lr = 0.001$. Optimization used Adam with $\texttt{ReduceLROnPlateau}$ on the validation trajectory loss; a fixed epoch budget ($E_{stop}=900$) is used with no early stopping, and the best checkpoint is selected post hoc. The batch size was 2500 samples with 100 batches per epoch. New trajectories were generated every 5 epochs, such that over 900 epochs the effective training dataset was 45M trajectories.

\paragraph{Results} The trajectory loss $\mathcal{L}_{\mathrm{traj}}$ and the total loss of Equation~\ref{eq:loss} are shown for the baseline model in Figure~\ref{fig:loss} in Appendix~\ref{app:training}. This shows that the spectral component learns the majority of the trajectory fit prior to the $\alpha$ ramp beginning. Ramping $\alpha$ enables the total loss to drop (through learning meaningful quantiles) and, after experiencing some transient instability, the loss converges steadily. The trajectory loss is the mean squared error (MSE). The final validation mean squared error averaged $(1.11\pm 0.13)\times 10^{-6}$ over different runs with different initial seeds. Notably, the spectral backbone alone (prior to $\alpha$ ramp-up) achieves an MSE less than $2\times$ the final MSE. 

\subsection{Conformal Quantile Regression}

We compare two models that are trained to predict the 95\%  Confidence Interval (CI) (2.5\%, 97.5\%) and 99\% CI (0.5\%, 99.5\%) quantiles. The CQR procedure was performed to best match the original predicted quantiles (95\% for the 95\% CI, 99\% for the 99\% CI). As shown in Table~\ref{tab:cqr}, models are well calibrated at the per-dimension, per-time prediction level. However, the 95\% CI quantile model fails to capture most \emph{full trajectories} at the 95\% level with a little more than 1\% of trajectories within the bounds. The same is true, but less dramatically, for the 99\% CI quantile model, which has 23\% of trajectories within the 99\% CI bounds. This justifies conformal scaling. Table~\ref{tab:cqr} demonstrates that the $\omega$ scaling is supportive of tighter bounds in the translational position and velocity dimensions. The results suggest that predicting a higher quantile may improve accuracy and reduce the need for CQR adjustments. However, such quantiles may be difficult to train in practice due to the low density of samples in the tail of the distribution. 

\begin{table}
\centering
\caption{Conformal quantile regression on the baseline trajectory model ($\ell{=}128$, $m{=}251$, seed 42), trained at two coverage targets. Raw cell-level coverage is close to the trained nominal in both cases, but raw trajectory-level (joint) coverage is far below the target. After conformal correction, the trajectory-level coverage reaches the target, with mean band width roughly $2{\times}$ the raw band width. Per-dimension weights $\omega[d]$ reflect difficulty in $R$-normalized units: body rates ($p, q$) and angles ($\phi, \theta$) are hardest, while translational positions ($x, y$) are easiest.}
\label{tab:cqr}
\vspace{0.2in}
\setlength{\tabcolsep}{5pt}
\begin{tabular}{lccccccc}
\hline
\textbf{Target} & \multicolumn{2}{c}{\textbf{Raw coverage}} & \multicolumn{2}{c}{\textbf{Post-CQR coverage}} & \textbf{Raw width} & \textbf{Post-CQR width} & \textbf{$Q_\eta$} \\
 & cell & joint & cell & joint & (mean) & (mean) & \\
\hline
95\% & 94.94\% & 1.10\%  & 99.95\% & 95.31\% & 0.00327 & 0.00775 & 0.00093 \\
99\% & 99.05\% & 22.64\% & 99.99\% & 99.00\% & 0.00433 & 0.00855 & 0.00087 \\
\hline
\end{tabular}

\vspace{0.5em}

\begin{tabular}{lcccccccccccc}
\hline
\textbf{Target} & $u$ & $x$ & $v$ & $y$ & $w$ & $z$ & $p$ & $\phi$ & $q$ & $\theta$ & $r$ & $\psi$ \\
\hline
95\% & 0.874 & 0.244 & 0.877 & 0.231 & 1.396 & 0.398 & 2.819 & 2.771 & 2.694 & 2.763 & 0.453 & 1.584 \\
99\% & 0.883 & 0.248 & 0.901 & 0.240 & 1.423 & 0.415 & 2.840 & 2.841 & 2.703 & 2.830 & 0.384 & 1.509 \\
\hline
\end{tabular}
\end{table}

\begin{figure}[ht]
    \centering
    \includegraphics[width=0.98\linewidth, trim={0 147cm 0cm 0cm}, clip]{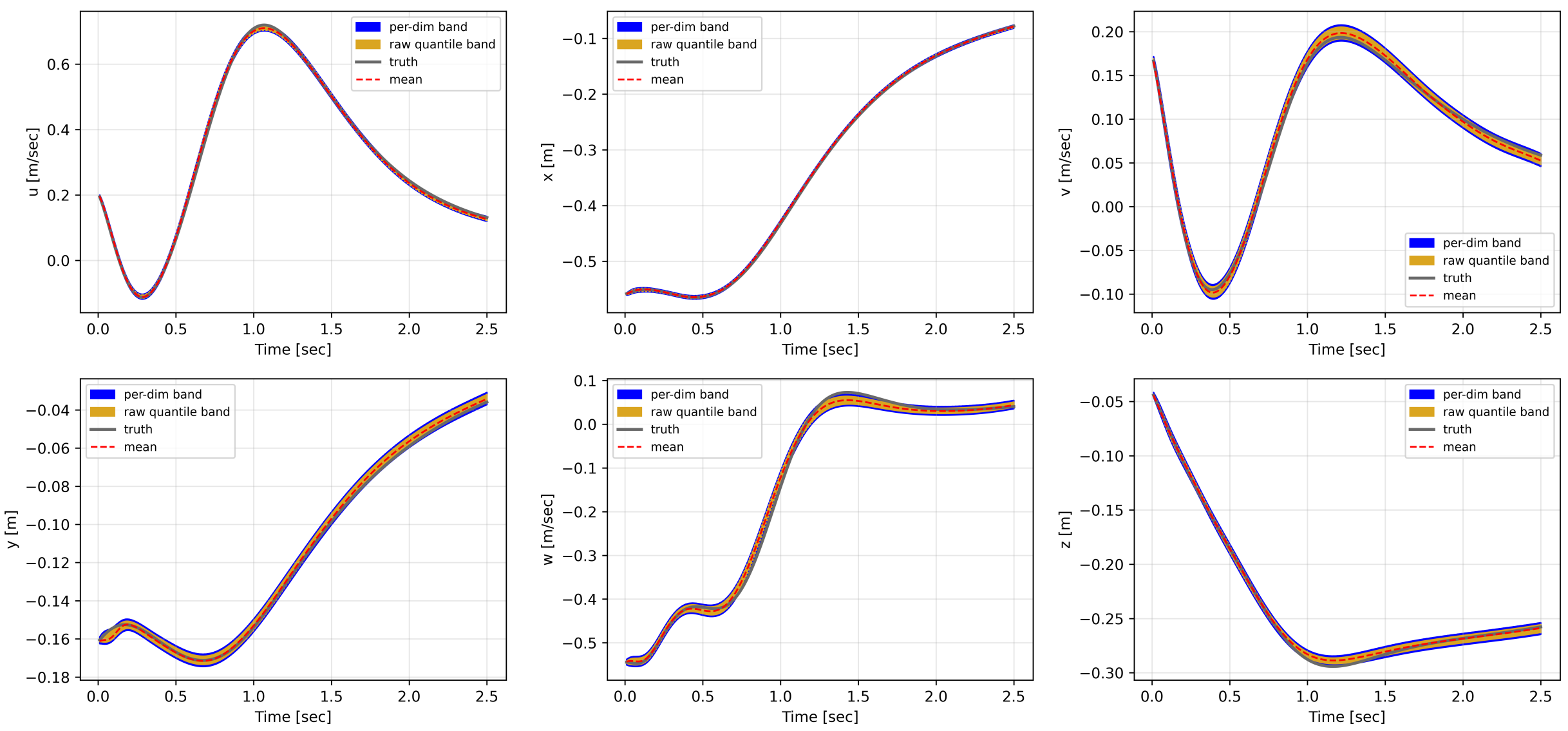}
    \caption{Conformal-calibrated trajectory bands for the \emph{worst-case hold-out trajectory prior to conformal calibration} under the 95\% quantile model (baseline: $\ell{=}128$, $m{=}251$, seed 42). Solid gray is the true trajectory; red dashed is the model's mean prediction; gold is the raw quantile band before calibration; blue is the per-dim conformal-calibrated band at 95\% joint coverage. The "worst case" is the test trajectory whose raw band has the largest miss before calibration; the true trajectory is contained at every $(t, d)$.}
\label{fig:example_95}
\end{figure}

At both targets, calibration roughly doubles the band width (Table~\ref{tab:cqr}); this is the price of joint coverage, and its practical acceptability is application-dependent. In Figure \ref{fig:example_95}, the hold-out sample that performed worst prior to CQR adjustments is shown (first six translational dimensions --- $(u,x,v,y,w,z)$) with the CQR adjustments applied to the quantile band for the 95\% CI baseline trained model. All dimensions and an example at the $99\%$ CI are provided in Appendix~\ref{app:full_figures}. The tightness of the interval and the value of the $Q_\eta$ and $\vec{R} \odot \vec{\omega}$ weights are clear with the true trajectory falling within the 95\% interval.

Having validated band-level joint coverage above, we statistically assess the additional margin that the true trajectory's own spline-approximation gap requires beyond the calibrated band by comparing post-CQR widths to the residual from Eq.~\ref{eq:deboor_bound} across 5000 test trajectories. The results in Table~\ref{tab:eps-knot-vs-half-width} compare (1) the per-trajectory maximum residual between the true trajectory and the least-squares (spline) fit (mean and 95\textsuperscript{th} percentile) with (2) the neural-network-and-CQR derived widths. These comparisons show that the additional margin is comparatively small for translational quantities (<10\%) and for rotational angles ($\phi/\theta$, $\psi$: 5-16\%), and larger (30-45\%) for rotational rates ($p/q$, $r$) with $r$ having the most margin to add when extending the guarantee to the true (non-idealized) trajectory. This may be caused by the particularly narrow half-width in $r$ (as it is usually well contained). The control point interval margin is wider than the trajectory half-width because the control point intervals have fewer assumptions on the underlying functions. 

\begin{table}[tb]
    \centering
    \caption{The sup-norm (per-trajectory maximum) residual between the true trajectory and its own least-squares spline fit, summarized as a mean ($\epsilon_{\text{knot}}$)  and 95th percentile ($\epsilon_{\text{knot}}^{p95}$) across trajectories, compared against the calibrated trajectory half-width ($\text{Traj}_{hw}$). The ratio is $\epsilon_{\text{knot}}^{p95}/\text{Traj}_{hw}$. Results from a 5000-trajectory test run. Some channels (e.g., $u$ and $v$) are combined due to the similarity of their values.}
    \label{tab:eps-knot-vs-half-width}
    \begin{tabular}{l|cccc|cccc}
        \hline
         & $u/v$ & $x/y$ & $w$ & $z$ & $p/q$ & $\phi/\theta$  & $r$ & $\psi$ \\
         & [m/s] & [m]  & [m/s] & [m] & [rad/s] & [rad] & [rad/s] & [rad] \\
        \hline
        $\epsilon_{\text{knot}}$ ($\times10^{-4}$)& 5.23 & 1.36 & 7.23 & 1.39 & 14.8 & 2.36 & 1.80 & 0.19 \\
        $\epsilon_{\text{knot}}^{p95}$ ($\times10^{-3}$)    & 1.07 & 0.28 & 1.41 & 0.28 & 3.75 & 0.46 & 0.70 & 0.07 \\
        Traj.$_{hw}$ ($\times10^{-2}$) & 1.28 & 1.01 & 2.06 & 1.65 & 1.11 & 0.28 & 0.16 & 0.15 \\
        $\epsilon_{\text{knot}}^{p95}/\text{Traj}_{hw}$ (\%)& 8 & 3 & 7 & 2 & 34 & 16  & \textbf{43} & 5 \\
        \hline
    \end{tabular}
\end{table}

\subsection{Ablation Studies}
We performed ablation studies, focusing on the effect of rollout size, encoder size, and latent lift dimensions on the fit through the total and validation trajectory loss. The lift dimension controls how much information can be passed from the encoder to the spectral layer. For this reason, we expect that the lift dimension (last encoder dimension) is the most impactful dimension for the encoder. The full results are in Appendix~\ref{app:ablation}. We find that the lift dimension is the most influential parameter within the network. 

\subsection{Knot Placement and Spectral Backbone Impact}
In addition to the above ablation studies, we ran separate studies where we replaced the adaptive knot placement with uniform knot placement and the spectral backbone with a recurrent neural network (RNN) backbone to study the importance of these aspects to the overall solution. The results are documented in Appendix~\ref{app:impact_knots} and Appendix~\ref{app:rnn}, respectively. The uniform knot placement performs worse than adaptive knots but demonstrates that this approach works with different knot placement schemes. The spectral backbone version converges to a lower loss and has tighter CQR bands than the RNN backbone. In addition, an RNN is slower at inference and cannot avail itself of the "tricks" described in Section~\ref{sec:hardware}. 

\subsection{Initial Condition Uncertainty Propagation}
\label{sec:ic_uncertainty}
The conformal analysis performed above relies on the exchangeability of the data. We implicitly assumed that the inference distribution is the same as the training distribution. This assumption is certainly violated for real systems, which always have sensor-induced noise. \cite{tibshirani2019conformal} provides the requisite analysis to enable one to extend the method above under covariate shift between training and inference due to noise. 
However, such analysis relies on noise-induced distribution shifts that must be justified either through empirical data or a verified noise model --- which we currently do not have. Appendix~\ref{app:ICUP} shows how initial-condition noise yields probabilistic bands without coverage guarantees (e.g., when only the current point's uncertainty is known).

\subsection{Inference Speed on Constrained Hardware}
\label{sec:hardware}
We evaluated inference speeds on a Raspberry Pi Zero 2 W with a quad-core 64-bit ARM Cortex-A53 processor and 512~MB of SDRAM in order to mimic deployment on constrained edge hardware. We used ONNX version 1.12.0 (opset 17) for inference implementation. Because ONNX is unable to support complex numbers, the conversion to ONNX format necessitated the expansion of the spectral decomposition into real and imaginary components. We leveraged the linearity between the spectral layer and the projection and decoder operations by noting that using $c_j = \sum_{\ell'} {Q_\theta^H}_{j,\ell'} \vec\chi^{\ell}_0$, we can rewrite

\begin{align}
\Big(\Gamma^{T}\, Q_\theta\, e^{\Lambda_\theta \tau}\, Q_\theta^{H}\, \vec\chi_0\Big)_{d}
&= \sum_{\ell} \Gamma_{\ell d} \sum_{j} Q_\theta^{\ell j}\, e^{\lambda_j \tau} \underbrace{\sum_{\ell'} {Q_\theta^H}_{j \ell'}\, \vec\chi_0^{\ell'}}_{c_j} = \sum_{j} \underbrace{\Big[\textstyle\sum_{\ell} \Gamma_{\ell d}\, Q_\theta^{\ell j}\Big]}_{\Psi_{dj}}\; c_j\; e^{\lambda_j \tau},
\label{eq:reduced_size}
\end{align}

where $\Gamma_{\ell d}$ represents any of the linear projection and decoder operations (there are four versions of $\Gamma$, one for each of $P$, $D^{mean}_\theta\bar{P}$, $D^{lo}_\theta\bar{P}$, and $D^{hi}_\theta\bar{P}$; see Equations~\ref{eq:decoder12},~\ref{eq:decoder3}, and ~\ref{eq:decoder4}). Both $\Psi_{dj} = \sum_\ell\Gamma_{\ell d}  Q_{\theta}^{\ell j}$ and   $e^{\lambda_j\tau }$ can be precomputed. Because $\Psi_{dj}c_j$ is smaller than $c_je^{\lambda_j \tau}$ ( $n_{dim} \times \ell_\text{lift}$ versus $ m \times \ell_\text{lift}$ where $n_{dim}\sim 12$,  $m \sim 251$, and $\ell_\text{lift}$ is the lift dimension), computing $\Psi_{dj}$ first in the right-most expression of Equation~\ref{eq:reduced_size} reduces memory use on the memory-constrained Raspberry Pi. When combined with the precomputed matrices, this reordering reduced the mean run time from 24.84~ms to 3.38~ms — a roughly $7\times$ reduction in inference time. We ran 11,000 initial conditions through a pipeline that included the activity model used to identify knot locations, spline basis construction, trajectory prediction, and finally control point calculation via SciPy's optimized \texttt{make\_lsq\_spline} function. The data were processed as sequential samples (unbatched). The first 1000 samples were used as warm-up and discarded.  In Table~\ref{tab:rasp_speed}, we show the mean, median, and 95$\textsuperscript{th}$ percentile run times for each step and those values summed to an estimated total time.  We find that the total pipeline inference time is 9.81~ms per inference (95\% below 10~ms) with 3.45~ms on average coming from the trajectory prediction (the difference between 3.38~ms and 3.45~ms is experimental variation). This is comparable with controller speeds ($\sim 100~\text{Hz}$) for drones \citep{mandaokar2026distributionally}. 
Using the numerical integrator used in data generation for trajectory estimation, the trajectory estimation time increased to 16.10~ms (95\% below 21.65~ms), increasing the average total time to 22.46~ms. This is slower than our method. However, we note that this numerical integration is optimized for accuracy, not speed, and may not provide a fair comparison.
\begin{table}[h]
\centering
\caption{Per-stage inference latency on the Raspberry Pi Zero 2 W. The focus of this paper is the trajectory prediction component, part 3. 
The activity model is a method for estimating an optimal knot density that neither affects nor is affected by the method described in this paper \citep{ongoing}.}
\begin{tabular}{c|l|c|c|c}\hline
     Stage & Description & Mean (ms) & Median (ms) & 95th pct (ms)\\\hline
     1 & Activity Model (optional to get knot locations) & 2.07 & 2.04 & 2.12 \\
     2 & Spline Basis Construction & 1.48 & 1.46 & 1.55 \\
     3 & \textbf{Trajectory Prediction (ML - This Paper)} & \textbf{3.45} & \textbf{3.41} & \textbf{3.49} \\
     4 & Control-Point Calculation & 2.81 & 2.77 & 2.97 \\ \hline
     \multicolumn{2}{c|}{\textbf{Total with ML (1+2+3+4)}} & \textbf{9.81} & \textbf{9.69} & \textbf{10.00} \\
\end{tabular}
\label{tab:rasp_speed}
\end{table}

\subsection{Control-Point Intervals with Coverage Guarantees} 
\label{sec:cp_conf}

\begin{figure}
    \centering
    \includegraphics[width=0.98\linewidth, trim={0 147cm 0cm 0cm}, clip]{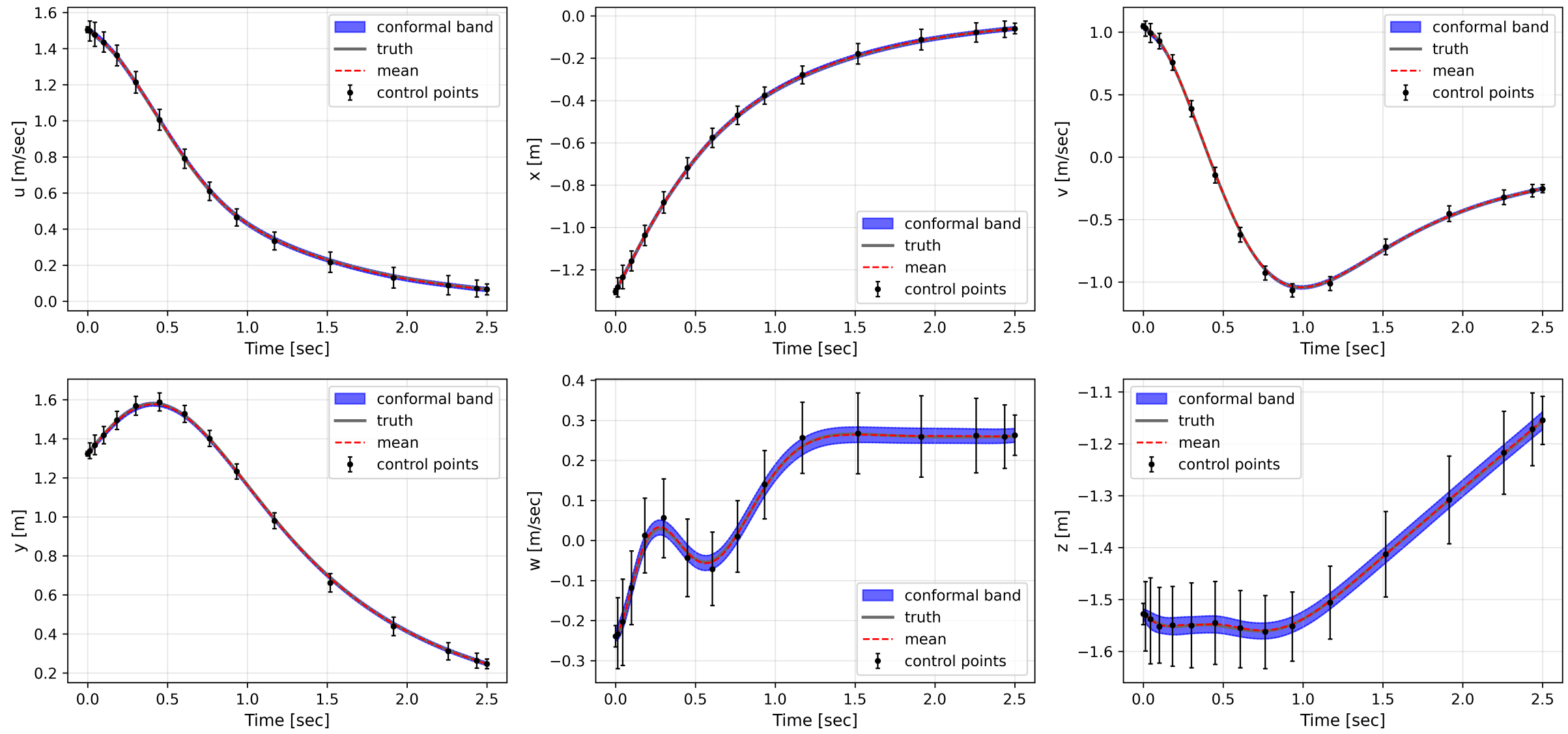}
    \caption{Calibrated trajectory bands and control-point intervals for one test trajectory, across six of twelve state dimensions $(u,x, v, y, w,z)$ ($\ell{=}128$, $m{=}251$, 95\% joint coverage). Gray solid is the true trajectory; red dashed is the mean prediction; the blue band is the per-dim conformal-calibrated band; black dots with vertical bars show the derived control points ($c_{i,\text{mid}}^d$) and their intervals $[c_{i,\text{lo}}^d, c_{i,\text{hi}}^d]$ at the Greville abscissae of the knot vector. The control-point intervals contain the spline coefficients of any band-respecting trajectory.}
\label{fig:control_points}
\end{figure}

An example trajectory with computed control points and their intervals (shown as error bars) is shown in Figure~\ref{fig:control_points}. The time to compute these bounds on the control points is negligible compared with the trajectory prediction and control point calculation. The linear relationship between control point intervals and the width of the conformal band is visible across the six translational dimensions (see Appendix~\ref{app:full_figures} for the full 12 dimensions). The overall tightness of the control point intervals demonstrates the value of this approach.

\section{Discussion}

By combining spectral-based trajectory prediction, conformal uncertainty quantification, and control-point derivation, the proposed architecture produces both a trajectory prediction and an associated finite-dimensional uncertainty representation. 
Trajectory loss improvement (Figure~\ref{fig:loss}) is essentially unchanged before and after the decoders are introduced, aside from brief warm-up instability. This is consistent with the spectral layer on a lifted space driving the trajectory fit, with the decoders contributing corrections and the quantile bands.


A key motivator was the application of this representation in safety assessment for autonomous systems. As shown in Section~\ref{sec:method_extend_quantiles}, convex constraints can be evaluated directly on the control-point intervals — conservatively, and without numerical integration. This supports an admissibility test by confidence interval that can be used to enable probabilistic safety filters that determine whether a desired reference is admissible. Our compact, uncertainty-aware description of the trajectory may be more suitable for decision-making algorithms than pointwise trajectory representations. 


\subsection{Future Work} Several directions that merit future study are identified below.

\textbf{Other Dynamical Systems and Hardware} 
A natural next step is to demonstrate this method on other autonomous ODEs such as the Duffing oscillator and ODEs of increasing complexity. Relaxing constraints on stability and fixed points and testing on systems such as the Lorenz attractor will stress-test our assumptions and the usefulness of the coverage bands. In addition, our hardware benchmark is limited to a single system. Future work may include additional physical hardware.

\textbf{Conformal Prediction under Covariate Shift} The initial-condition noise propagation of Section~\ref{sec:ic_uncertainty} provides probabilistic bands but no coverage guarantee. Extending the conformal procedure to sensor-noise covariate shift via the weighting of \citet{tibshirani2019conformal} could restore guarantees, at the cost of requiring an empirically justified or verified noise model.

\textbf{Smoothness-Informed Control Points} Section~\ref{sec:method_extend_quantiles} suggests a path to tighten control-point intervals while preserving coverage by restricting the admissible set to dynamically valid functions. Implementing and benchmarking the resulting linear programs is a natural next step. 

\textbf{Safety Filters and Reference Repair} The admissibility conditions of Section~\ref{sec:method_extend_quantiles} determine whether a desired reference is safe. Extending them to \emph{repair} an inadmissible reference requires searching over candidate references and a formalization of the constraint sets. This is a direct path to a deployable safety filter.


\section{Conclusion}

This work demonstrated an end-to-end neural-network-based method to map initial conditions to control-point intervals that contain the least-squares fit of the true trajectory for a quadrotor under a closed-loop LQR with at least $(1-\delta)$ coverage under the standard exchangeability assumption of conformal prediction. 

We combined and modified several established procedures.  First, we trained a neural network with a spectral layer to provide the backbone expected-value point prediction. Three decoder heads refined the point estimate and generated quantile estimates. Then, we used conformal quantile regression with a supremum across dimensions and time steps, similar to \citet{diquigiovanni2022conformal}, in order to provide joint conformal coverage. Finally, the resulting bands are mapped to spline control points, giving a compact, uncertainty-aware representation.

To our knowledge, the application of these techniques together is new. We benchmarked the performance on a Raspberry Pi Zero 2 W and showed that after refactoring for inference the trajectory model 
can support control processes at operation speeds of 100~Hz. 
Because this method relies on general properties of ODEs with stable fixed points rather than quadrotor-specific dynamics, we anticipate that it may generalize to other ODE systems, which we leave to future work. 



\section*{Acknowledgements}
Copyright 2026 Carnegie Mellon University and Duquesne University

This material is based upon work supported by the Department of War under Air Force Contract No. FA8702-15-D-0002 with Carnegie Mellon University for the operation of the Software Engineering Institute, a federally funded research and development center.  

References herein to any specific entity, product, process, or service by trade name, trademark, manufacturer, or otherwise, does not necessarily constitute or imply its endorsement, recommendation, or favoring by Carnegie Mellon University or its Software Engineering Institute nor of Carnegie Mellon University - Software Engineering Institute by any such named or represented entity.

[DISTRIBUTION STATEMENT A] This material has been approved for public release and unlimited distribution.  Please see Copyright notice for non-US Government use and distribution. DM26-0292

\bibliographystyle{tmlr} 
\bibliography{references}

@inproceedings{giorgiani2020quadcopter, 
title={Quadcopter Control Optimization through Machine Learning}, 

DOI={10.2514/6.2020-1148}, 
booktitle={AIAA Scitech 2020 Forum}, 
publisher={American Institute of Aeronautics and Astronautics}, 
author={Giorgiani do Nascimento, Renato and Fricke, Kajetan and Viana, Felipe}, year={2020}, 
month=Jan }

@ARTICLE{safe_fast_mpc_nn_2020,
  author={Nubert, Julian and Köhler, Johannes and Berenz, Vincent and Allgöwer, Frank and Trimpe, Sebastian},
  journal={IEEE Robotics and Automation Letters}, 
  title={Safe and {F}ast {T}racking on a {R}obot {M}anipulator: {R}obust {MPC} and {N}eural {N}etwork {C}ontrol}, 
  year={2020}, 
  month=Apr, 
  volume={5},
  number={2},
  pages={3050-3057},
  doi={10.1109/LRA.2020.2975727},
  }

@article{dietterich2022conformal,
  title={Conformal prediction intervals for {M}arkov {D}ecision {P}rocess trajectories},
  author={Dietterich, Thomas G. and Hostetler, Jesse},
  journal={arXiv preprint arXiv:2206.04860},
  year={2022}
}

@inproceedings{bansal2017hamilton,
  title     = {Hamilton-{J}acobi reachability: A brief overview and recent advances},
  author    = {Bansal, Somil and Chen, Mo and Herbert, Sylvia and Tomlin, Claire J.},
  booktitle = {2017 IEEE 56th Annual Conference on Decision and Control (CDC)},
  pages     = {2242--2253},
  year      = {2017},
  organization = {IEEE}
}

@article{garone2017reference,
title = {Reference and command governors for systems with constraints: A survey on theory and applications},
journal = {Automatica},
publisher={Elsevier}, 
volume = {75},
pages = {306-328},
year = {2017},
month=Jan,
DOI={10.1016/j.automatica.2016.08.013},
url = {https://www.sciencedirect.com/science/article/pii/S0005109816303715},
author = {Emanuele Garone and Stefano {Di Cairano} and Ilya Kolmanovsky},
}

@book{rawlings2017model,
  title     = {Model Predictive Control: Theory, Computation, and Design},
  author    = {Rawlings, James B. and Mayne, David Q. and Diehl, Moritz},
  edition   = {2},
  publisher = {Nob Hill Publishing},
  address   = {Madison, WI},
  year      = {2017}
}

@inproceedings{ongoing,
    author = {Ratchford, Jasmine and Romagnoli, Raffaele},
    title = {A Neural Network Approach towards Real-Time Knot Point Density Estimates for the Nonlinear Quadrotor},
    booktitle = {Proceedings of the 65th IEEE Conference on Decision and Control (CDC)},
    year      = {2026},
    address   = {Honolulu, Hawaii, USA},
    month     = dec,
    note={To appear}
}

@book{de1978practical,
  title={A practical guide to splines},
  author={De Boor, Carl},
  volume={27},
  year={1978},
  publisher={Springer New York}
}

@article{devore1998nonlinear,
  title={Nonlinear approximation},
  author={DeVore, Ronald A},
  journal={Acta Numerica},
  volume={7},
  pages={51--150},
  year={1998},
  publisher={Cambridge University Press}
}

@Article{van2015b,
AUTHOR = {Van Loock, Wannes and Pipeleers, Goele and Swevers, Jan},
TITLE = {B-spline parameterized optimal motion trajectories for robotic systems with guaranteed constraint satisfaction},
JOURNAL = {Mechanical Sciences},
VOLUME = {6},
YEAR = {2015},
NUMBER = {2},
PAGES = {163--171},
URL = {https://ms.copernicus.org/articles/6/163/2015/},
DOI = {10.5194/ms-6-163-2015}
}

@article{brunton2016koopman,
  title={Koopman invariant subspaces and finite linear representations of nonlinear dynamical systems for control},
  author={Brunton, Steven L. and Brunton, Bingni W. and Proctor, Joshua L. and Kutz, J. Nathan},
  journal={PLoS ONE},
  volume={11},
  number={2},
  pages={e0150171},
  year={2016},
  publisher={Public Library of Science San Francisco, CA USA},
  notes={Koopman conjugacy; Need for singluar fixed point}
}

@article{lusch2018deep, 
title={Deep learning for universal linear embeddings of nonlinear dynamics}, 
volume={9}, 
DOI={10.1038/s41467-018-07210-0}, 
number={1}, 
journal={Nature Communications}, publisher={Springer Science and Business Media LLC}, author={Lusch, Bethany and Kutz, J. Nathan and Brunton, Steven L.}, year={2018}, month=Nov }

@inproceedings{tibshirani2019conformal,
 author = {Tibshirani, Ryan J. and Foygel Barber, Rina and Cand{\`e}s, Emmanuel and Ramdas, Aaditya},
 booktitle = {Advances in Neural Information Processing Systems},
 editor = {H. Wallach and H. Larochelle and A. Beygelzimer and F. d\textquotesingle Alch\'{e}-Buc and E. Fox and R. Garnett},
 pages = {},
 publisher = {Curran Associates, Inc.},
 title = {Conformal {P}rediction {U}nder {C}ovariate {S}hift},
 url = {https://proceedings.neurips.cc/paper_files/paper/2019/file/8fb21ee7a2207526da55a679f0332de2-Paper.pdf},
 volume = {32},
 year = {2019}
}

@article{rousseau2019minimum,
  author  = {Rousseau, Gauthier and {Stoica Maniu}, Cristina and
             Tebbani, Sihem and Babel, Mathieu and Martin, Nicolas},
  title   = {Minimum-time {B}-spline trajectories with corridor constraints.
             {A}pplication to cinematographic quadrotor flight plans},
  journal = {Control Engineering Practice},
  publisher={Elsevier},
  volume  = {89},
  pages   = {190--203},
  year    = {2019},
  doi     = {10.1016/j.conengprac.2019.05.022}
}

@inproceedings{romano2019conformalized,
 author = {Romano, Yaniv and Patterson, Evan and Cand{\`e}s, Emmanuel},
 booktitle = {Advances in Neural Information Processing Systems},
 editor = {H. Wallach and H. Larochelle and A. Beygelzimer and F. d\textquotesingle Alch\'{e}-Buc and E. Fox and R. Garnett},
 pages = {},
 publisher = {Curran Associates, Inc.},
 title = {Conformalized Quantile Regression},
 url = {https://proceedings.neurips.cc/paper_files/paper/2019/file/5103c3584b063c431bd1268e9b5e76fb-Paper.pdf},
 volume = {32},
 year = {2019},
 numpages = {11}
}

@article{pan2020physics,
author = {Pan, Shaowu and Duraisamy, Karthik},
title = {Physics-{I}nformed {P}robabilistic {L}earning of {L}inear {E}mbeddings of {N}onlinear {D}ynamics with {G}uaranteed {S}tability},
journal = {SIAM Journal on Applied Dynamical Systems},
volume = {19},
number = {1},
pages = {480-509},
year = {2020},
doi = {10.1137/19M1267246},
eprint = { 
        https://doi.org/10.1137/19M1267246
}
}

@inproceedings{stankeviciute2021conformal,
 author = {Stankevi{\v c}i{\=u}t{\.e}, Kamile and M. Alaa, Ahmed and van der Schaar, Mihaela},
 booktitle = {Advances in Neural Information Processing Systems},
 editor = {M. Ranzato and A. Beygelzimer and Y. Dauphin and P.S. Liang and J. Wortman Vaughan},
 publisher = {Curran Associates, Inc.},
 title = {Conformal Time-series Forecasting},
 url = {https://proceedings.neurips.cc/paper_files/paper/2021/file/312f1ba2a72318edaaa995a67835fad5-Paper.pdf},
 volume = {34},
 year = {2021}
}

@article{brunton2022modern,
author = {Brunton, Steven L. and Budi{\v s}i{\'c}, Marko and Kaiser, Eurika and Kutz, J. Nathan},
title = {Modern {K}oopman {T}heory for {D}ynamical {S}ystems},
journal = {SIAM Review},
volume = {64},
number = {2},
pages = {229-340},
year = {2022},
doi = {10.1137/21M1401243},
eprint = {https://doi.org/10.1137/21M1401243}
}

@article{otto2019linearly,
author = {Otto, Samuel E. and Rowley, Clarence W.},
title = {{L}inearly {R}ecurrent {A}utoencoder {N}etworks for {L}earning {D}ynamics},
journal = {SIAM Journal on Applied Dynamical Systems},
volume = {18},
number = {1},
pages = {558-593},
year = {2019},
doi = {10.1137/18M1177846},
eprint = { 
        https://doi.org/10.1137/18M1177846
}
}

@article{otto2021koopman,
  title   = {Koopman {O}perators for {E}stimation and {C}ontrol of {D}ynamical {S}ystems},
  author  = {Otto, Samuel E. and Rowley, Clarence W.},
  journal={Annual Review of Control, Robotics, and Autonomous Systems},
  volume  = {4},
  pages   = {59--87},
  year    = {2021},
  month   = May,
  publisher = {Annual Reviews},
  doi     = {10.1146/annurev-control-071020-010108},
}

@article{diquigiovanni2022conformal,
author={Diquigiovanni, Jacopo and Fontana, Matteo and Vantini, Simone},
title={Conformal prediction bands for multivariate functional data},  
journal={Journal of Multivariate Analysis},
volume={189}, 
DOI={10.1016/j.jmva.2021.104879}, 
publisher={Elsevier}, 
year={2022}, 
month=May, pages={104879} }

@book{moore2009introduction,
title={Introduction to Interval Analysis}, 
ISBN={9780898717716}, DOI={10.1137/1.9780898717716}, publisher={Society for Industrial and Applied Mathematics}, author={Moore, Ramon E. and Kearfott, R. Baker and Cloud, Michael J.}, year={2009}, month=Jan }

@book{neumaier1990interval,
  author    = {Neumaier, Arnold},
  title     = {Interval Methods for Systems of Equations},
  publisher = {Cambridge University Press},
  series    = {Encyclopedia of Mathematics and its Applications},
  volume    = {37},
  year      = {1990}
}

@inproceedings{ICLR2024_sun_copula,
 author = {Sun, Sophia and Yu, Rose},
 booktitle = {International Conference on Learning Representations},
 editor = {B. Kim and Y. Yue and S. Chaudhuri and K. Fragkiadaki and M. Khan and Y. Sun},
 pages = {31819--31840},
 title = {Copula Conformal prediction for multi-step time series prediction},
 url = {https://proceedings.iclr.cc/paper_files/paper/2024/file/8707924df5e207fa496f729f49069446-Paper-Conference.pdf},
 year = {2024}
}

@article{jamil2026lie,
  title={Lie Generator Networks for Nonlinear Partial Differential Equations},
  author={Jamil, Shafayeth and Kapadia, Rehan},
  journal={arXiv preprint arXiv:2603.29264},
  year={2026}
}

@article{mandaokar2026distributionally,
  author={Mandaokar, Dnyandeep and Rinner, Bernhard},
  journal={IEEE Robotics and Automation Letters}, 
  title={Distributionally {R}obust {A}cceleration {C}ontrol {B}arrier {F}ilter for {E}fficient {UAV} {O}bstacle {A}voidance}, 
  year={2026},
  publisher={Institute of Electrical and Electronics Engineers (IEEE)},
  month=May,
  volume={11},
  number={5},
  pages={5914-5921},
  doi={10.1109/LRA.2026.3677752}
  }

@article{wabersich2023data,
  title={Data-driven safety filters: {H}amilton-{J}acobi reachability, control barrier functions, and predictive methods for uncertain systems},
  author={Wabersich, Kim P. and Taylor, Andrew J. and Choi, Jason J. and Sreenath, Koushil and Tomlin, Claire J. and Ames, Aaron D. and Zeilinger, Melanie N.},
  journal={IEEE Control Systems Magazine},
  volume={43},
  number={5},
  pages={137--177},
  year={2023},
  publisher={IEEE}
}

\newpage
\appendix
\section{Quadrotor Dynamics and Data-Generation Parameters}
\label{app:dynamics}
We consider a six-degree-of-freedom rigid-body model of a quadrotor. The system can be described with the following equations:
\begin{align}
    \dot{u} &= rv - qw - g\sin\theta \\
    \dot{v} &= pw - ru + g\cos\theta\sin\phi \\
    \dot{w} &= qu - pv + g\cos\theta\cos\phi - F_1/m_{kg} \\
    \dot{p} &= \frac{(J_y - J_z)}{J_x}qr + F_2/J_x \\
    \dot{q} &= \frac{(J_z - J_x)}{J_y}pr + F_3/J_y \\
    \dot{r} &= \frac{(J_x - J_y)}{J_z}pq + F_4/J_z
\end{align}
where $(u,v,w)$ are the translational velocities; $(p,q,r)$ are the rotational velocities; $m_{kg}$ is the mass; $(J_x, J_y, J_z)$ are the principal moments of inertia; $g$ is the acceleration due to gravity; $F_{1}$ is the total thrust force generated by the rotors; $(F_2, F_3, F_4)$ are the rolling, pitching, and yawing moment thrusts. These are combined with the six integral equations, where we use a small-angle approximation for the kinematic mapping and the closed-loop law:
\begin{equation}
    \mathbf{F} = \begin{bmatrix} F_f \\ 0 \\ 0 \\ 0 \end{bmatrix} - K(\mathbf{x} - \mathbf{x}_{sp})
\end{equation}
where $F_f$ is the trim thrust, $K$ the LQR gain matrix, and $\mathbf{x}_{sp}$ the setpoint state. The full values for the quadrotor system are given below in Table~\ref{tab:dynamics}.

\begin{table}[ht]
    \centering
    \begin{tabular}{rl}
        \hline
        \textbf{Parameter}& \textbf{Value}\\
        \hline
        \multicolumn{2}{c}{\textit{Quadrotor Dynamics }}\\
        \hline
        Total system mass              & 0.600 kg\\
        Quadrotor body mass            & 0.410 kg\\
        Propeller radius               & 0.0504 meters\\
        Arm length                     & 0.108 meters\\
        Propeller mass                 & 0.00311 kg\\
        Motor + propeller mass         & 0.0391 kg\vspace{0.1cm}\\
        
        \hline
        \multicolumn{2}{c}{\textit{Initial Conditions Range (maximum magnitudes)}}\\
        \hline
        $x^{ic}_{\max}$& 5.0 meters\\
        $\dot{x}^{ic}_{\max}$& 3.0 meters / second\\
        $\theta^{ic}_{\max}$& $\pi/4$ radians\\
        $\dot{\theta}^{ic}_{\max}$& $\pi/2$ radians / second\vspace{0.1cm}\\
        \hline
        \multicolumn{2}{c}{\textit{Discretization and Processing }}\\
        \hline
        $t_{samp}$&0.005 seconds\\
        T (total time) &2.5 seconds\\
    \end{tabular}
    \vspace{0.05cm }
    \caption{Parameter values used in data generation and processing. Table repeated from \cite{ongoing}.}
    \label{tab:dynamics}
\end{table}

\clearpage
\section{Architecture and Training Details}
\label{app:training}
\begin{figure}[h]
    \centering
    \includegraphics[width=0.8\linewidth]{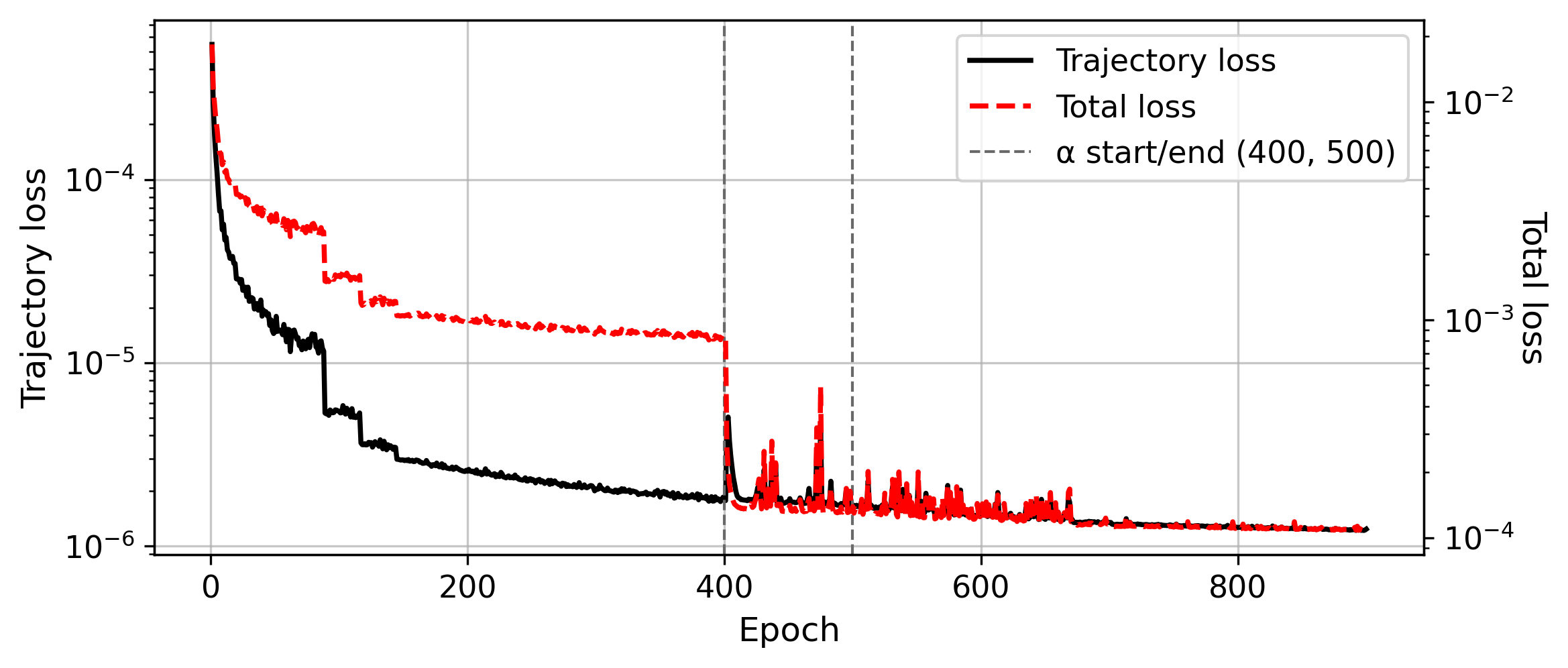}
    \caption{Validation loss curves for the baseline model ($\ell{=}128$, $m{=}251$): trajectory loss (mean MSE, black, left axis) and total loss including pinball band terms (red dotted, right axis). Vertical lines mark the $\alpha$ warmup window (epochs 400-500), during which band decoders activate and the loss briefly shows some instability before stabilizing and converging slowly. There are 62,500 samples in the validation dataset. These samples are generated once (unlike training data that is regenerated every 5 epochs) such that model performance is evaluated across the same data each epoch.}
\label{fig:loss}
\end{figure}

Here we present the details of the neural network architecture and an example of the training loss. Table~\ref{tab:traj_network} shows the sizes of each layer in the network. The spectral decomposition output is complex-valued and is equivalent to two $128$-sized real-valued tensors, thus giving the shape $(2,\, 128)$.  The loss equations for the total loss and trajectory loss shown in Figure~\ref{fig:loss} are given by Equation~\ref{eq:loss} and Equation~\ref{eq:spec_loss}, respectively. Excepting some training instability related to the warm-up period, the course of trajectory loss improvement is unchanged after $\alpha$ ramp-up. The total loss decreases precipitously because the upper and lower bounds are no longer fixed to the expected value.

\begin{table}[h]

    \caption{The baseline architecture used to predict the trajectory and quantile bands from initial conditions. The spectral decomposition evolves an $\ell{=}128$ lifted state; decoders read the $2\ell - d = 244$ non-state lifted channels and are bias-free.\vspace{0.1in}}
    \centering
        \begin{tabular}{lrcc}\hline
            \textbf{Layer Type} & \textbf{Input} & \textbf{Output} & \textbf{Param.}\\
            \hline
            Linear / SiLU & 12 & 128 & 1664\\
            Linear / SiLU & 128 & 64 & 8256\\
            Linear / SiLU & 64 & 128 & 8320\\
            Linear (lift) & 128 & 116 & 14964\\
            \hline
            Spectral decomposition & 128 & (2,\,128) & 33024\\
            \hline
            Decoder: mean (Linear) & 244 & 12 & 2928\\
            Decoder: low (Linear) & 244 & 12 & 2928\\
            Decoder: high (Linear) & 244 & 12 & 2928\\
            \hline
            \textbf{Total} & & & \textbf{75012}\\
    \end{tabular}\vspace{0.1cm}
    \label{tab:traj_network}
\end{table}

\clearpage

\section{Ablation Studies}
\label{app:ablation}
 Here we present results of an ablation study of our baseline network, varying the most impactful components of the \emph{lift} and \emph{evolve} framework. We find that the lift dimension is by far the most influential parameter. For a lift dimension of 64, the trajectory loss increased by about 50\% from $1.12\times10^{-6}$ to $1.69\times10^{-6}$. This occurs despite increasing the stopping epoch from $E_{stop}=900$ to $E_{stop}=1200$ because the models with lower lifted dimension are slower to converge after $\alpha$ is ramped to $1.0$. Increasing lifted dimension from 64 to 128 reduced trajectory loss by $\sim34\%$; further increasing from 128 to 256 reduced it by only $\sim 22\%$ despite adding roughly $4\times$ more spectral parameters ($\mathcal{O}(\ell^2)$ scaling). This suggests diminishing returns and motivates $\ell=128$ as a practical operating point. The encoder shape also plays a significant role in the accuracy of the model. However, increasing the size of all layers within the encoder is less efficient in terms of the total number of learned parameters. Integral consistency loss was largely ineffective, likely because such consistency was learned by the system and did not contribute significantly to the loss. Using a 99\% instead of 95\% quantile reduced the trajectory loss almost negligibly. The total loss for the 99\% quantile model is significantly lower. However, it is not directly comparable with the 95\% quantile bin loss. Rollout dimension had little effect on the overall loss.

\begin{table}[h]
    \centering
        \begin{tabular}{cc|S[table-format=1.3(2)]S[table-format=1.3(2)]}\hline
        
        \multicolumn{2}{c|}{Architecture} & \multicolumn{2}{c}{Validation Loss} \\ \hline
        Lifted & Rollout & {Trajectory Only} & {Total} \\
        Dim & Steps & {(mean MSE, $\times10^{-6}$)} & {(total, $\times10^{-4}$)} \\ \hline
        \multicolumn{4}{l}{\textit{Baseline (lifted = 128, rollout = 251, 3 seeds)}} \\
        128 & 251 & 1.12(13) & 1.050(94) \\ \hline
        \multicolumn{4}{l}{\textit{Lifted dimension sweep (rollout = 251, 3 seeds)}} \\
        64$^*$ & 251 & 1.69(19) & 1.144(64) \\
        256 & 251 & 0.875(53) & 0.841(22) \\ \hline
        \multicolumn{4}{l}{\textit{Encoder shape (lifted = 128, rollout = 251)}} \\
        \multicolumn{2}{c|}{$64 \times 64 \times 64$}  & 1.54 & 1.20 \\
        \multicolumn{2}{c|}{$128 \times 128 \times 128$}  & 0.654 & 0.71 \\
        \multicolumn{2}{c|}{$256 \times 256 \times 256$}  & 0.523 & 0.56 \\
        \multicolumn{2}{c|}{$128 \times 32 \times 128$}  & 1.47 & 1.15 \\ \hline
        \multicolumn{4}{l}{\textit{Rollout sweep (lifted = 128, 3 seeds)}} \\
        128 & 101 & 1.03(11) & 0.971(55) \\
        128 & 501 & 1.090(59) & 1.000(28) \\ \hline
        \multicolumn{4}{l}{\textit{Single Runs Tests (lifted = 128, rollout = 251, seed 42)}} \\
        \multicolumn{2}{l|}{Integral Consistency ($\lambda_{\text{int}}=0.25$)} & 1.24 & 1.09 \\
        \multicolumn{2}{l|}{99\% Quantile interval$^\dagger$} & 0.985 & 0.303 \\ \hline
        
        \end{tabular}

    \caption{Effect of lifted dimension, encoder shape, rollout length, physics consistency, and trained quantile interval on fit. Architecture rows report mean $\pm$ std over three seeds; ablation rows are single-seed. Trajectory Only is the MSE on the mean prediction (interval-invariant); Total is the total validation loss including pinball band terms. $^{*}$Trained for 1200 epochs rather than 900. $^\dagger$The validation loss at the 99\% interval is not directly comparable with the 95\% rows: the pinball weights $q$, $1{-}q$ are smaller at more extreme quantile levels.}
    \label{tab:arch_ablation}
\end{table}
\clearpage
\section{Full Twelve-Dimension Results}
\label{app:full_figures}

The following figures present examples of full results --- across all 12 dimensions for the model through its various stages of post-training conformal quantile regression and spline fitting. 

\begin{figure}[ht]
    \centering
    \includegraphics[width=0.98\linewidth ]{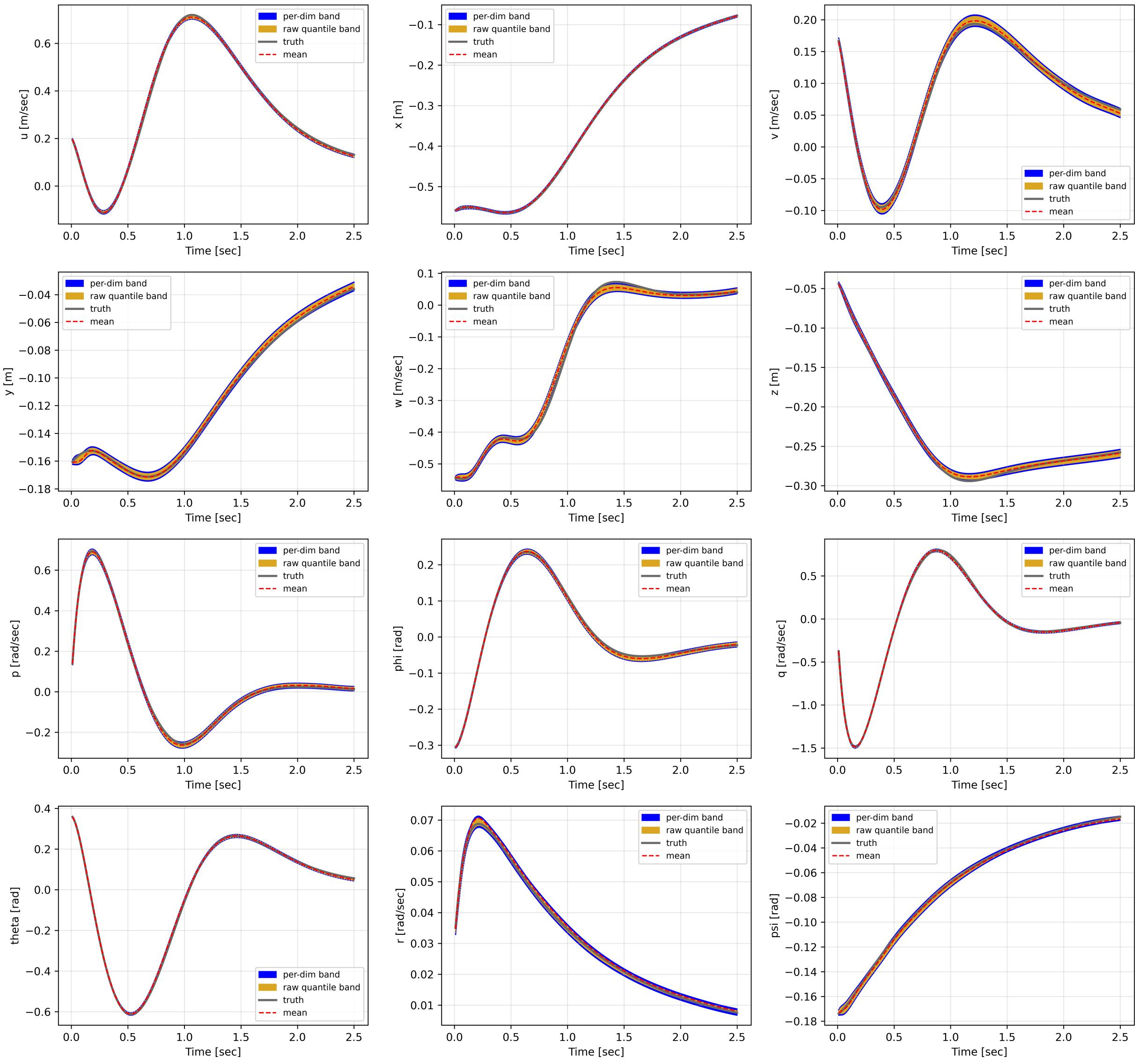} 
    \caption{Full twelve-dimension version of Figure~\ref{fig:example_95}: conformal-calibrated trajectory bands for the \emph{worst-case hold-out trajectory prior to conformal calibration} under the 95\% quantile model (baseline: $\ell{=}128$, $m{=}251$, seed 42). Solid gray is the true trajectory; red dashed is the model's mean prediction; gold is the raw quantile band before calibration; blue is the per-dim conformal-calibrated band at 95\% joint coverage.}
    \label{fig:cqr95_full}
\end{figure}

\begin{figure}[ht]
    \centering
    \includegraphics[width=0.98\linewidth]{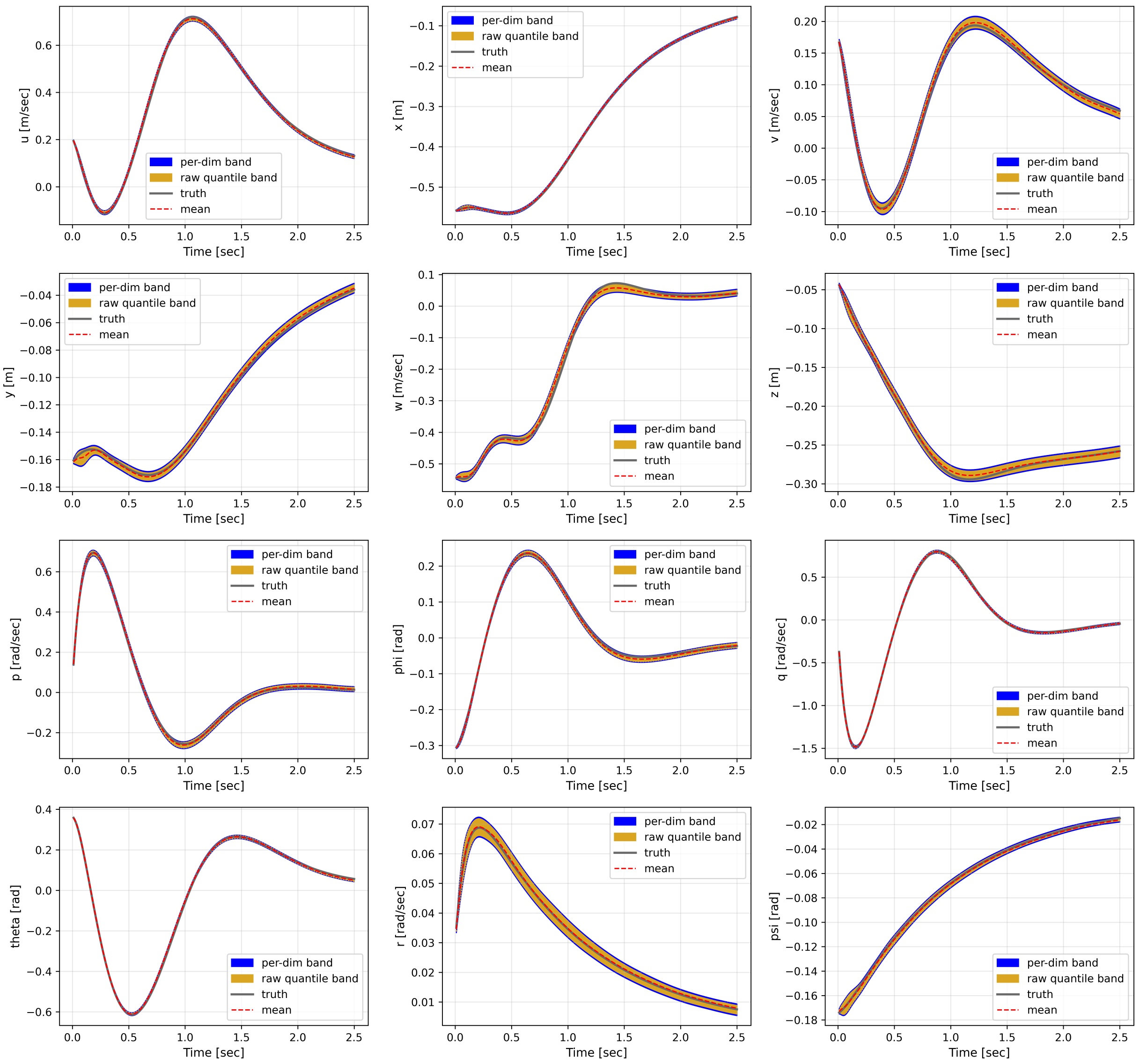} 
    \caption{Same as Figure~\ref{fig:cqr95_full}, but for the 99\% quantile model. The raw quantile bands (gold) are wider than at the 95\% target, and the conformal correction is correspondingly smaller; the calibrated band (blue) achieves 99\% joint coverage at a band width comparable with that of the 95\% case.}
    \label{fig:cqr99_full}
\end{figure}

\begin{figure}[ht]
    \centering
    \includegraphics[width=0.98\linewidth,]{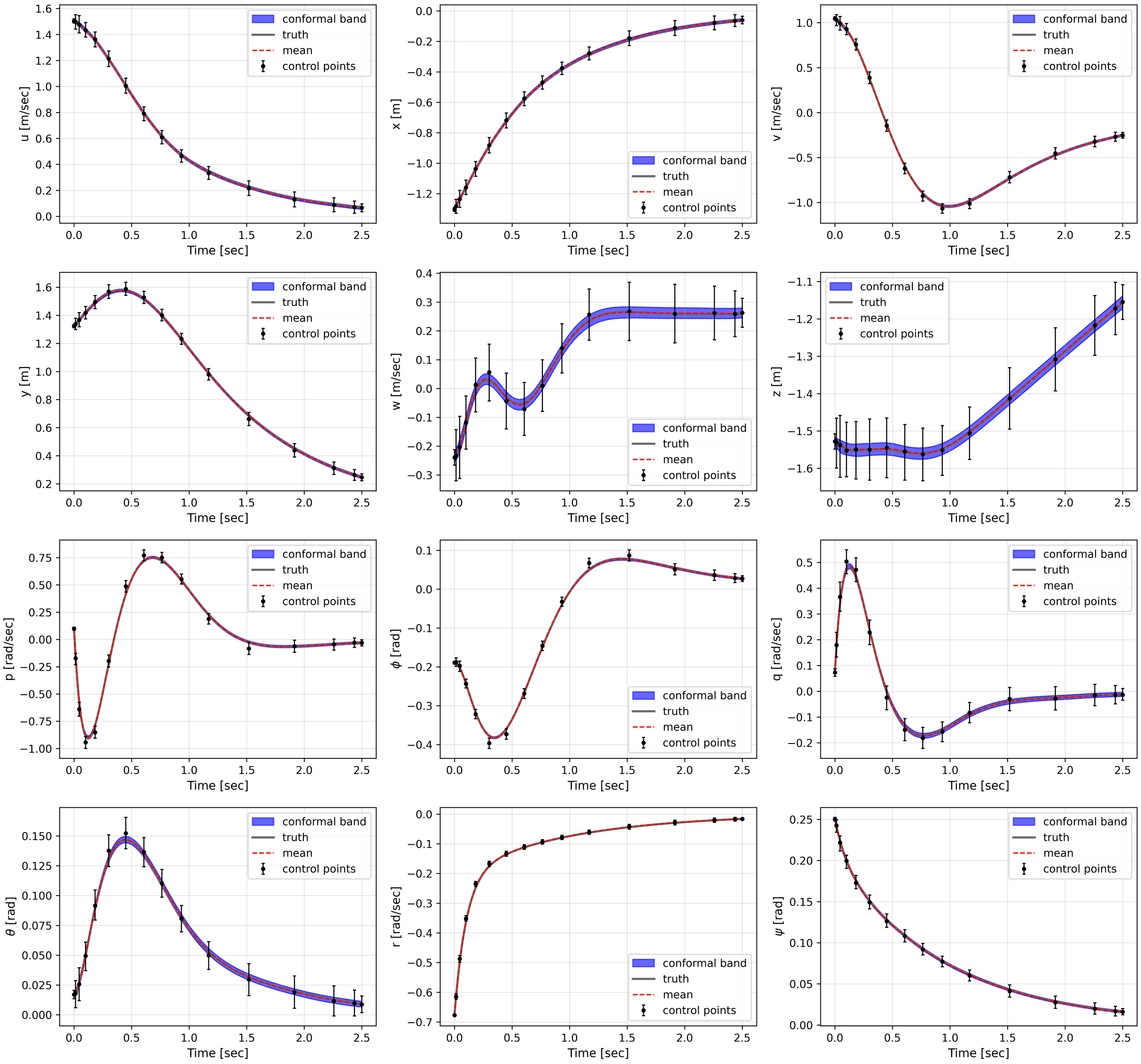}
    \caption{Full twelve-dimension version of
    Figure~\ref{fig:control_points}: calibrated trajectory bands and
    control-point intervals for one test trajectory ($\ell{=}128$,
    $m{=}251$, 95\% joint coverage). Gray solid is the true
    trajectory; red dashed is the mean prediction; the blue band is
    the per-dim conformal-calibrated band; black dots with vertical
    bars show the computed control points and their intervals at the
    Greville abscissae of the knot vector. There is no simulated noise on the initial condition. The generation process is identical to Figure~\ref{fig:cqr95_full} with the added computation of spline control points and their intervals, per Section~\ref{sec:cp_conf}.}
    \label{fig:control_points_full}
\end{figure}

\clearpage
\section{The Impact of Knot Point Placement}
\label{app:impact_knots}

Our previous work, \cite{ongoing}, parallels this paper, but focuses on adaptive knot placement motivated by the approximation bound in Equation~\ref{eq:deboor_bound}. Although we leverage that work to improve knot placement, the work presented in this paper can be used with any knot placement scheme, including uniform knot points. Here we describe the practical aspects of that paper in brief and show the impact of using uniform knots.

Starting from Equation~\ref{eq:deboor_bound}, we use the local quantities $h_i^2\|x''\|_{L^\infty([t_i,t_{i+1}])}$ as indicators of approximation difficulty, where $h_i$ is the spacing between consecutive knots $t_i$ and $t_{i+1}$. Under the assumption that the local curvature varies slowly within sufficiently fine knot spans, approximately balancing these indicators gives the knot-density relation
\[
\rho(t)\propto\sqrt{\|x''(t)\|}.
\]
This suggests that an effective knot-selection strategy should account for the local variation of the trajectory. The complete derivation and the associated local approximation assumption are presented in \cite{ongoing}.

For an autonomous ODE in which the initial condition $x_0$ determines the trajectory, the corresponding adaptive knot density can be estimated directly from $x_0$. We built a lightweight neural network with fully connected and convolutional layers that takes as input the same initial conditions $x_0$ and returns an estimate of $|x''|^{max}_{L^\infty([t_i,t_{i+1}])}$, where $^{max}$ signifies the local maximum $|x''|_{L^\infty([t_i,t_{i+1}])}$ across six of the twelve dimensions that define the quadrotor system (the other six being the corresponding positions). The full network size is approximately 22K parameters, meaning it can fit on resource-constrained hardware and run alongside the network described in this paper, as supported by the measured neural-network inference time. Using just the velocity states was a design choice; the same method could be applied to all twelve states. The network's output is converted to a knot density $\rho(t)=\sqrt{\hat{g}(t;x_0)/(8\varepsilon)}$ for a desired tolerance $\varepsilon$, with $\rho$ clipped to a minimum and maximum value to bound knot spacing. Knot locations are obtained by integrating $\rho(t)$ over time. The results from \cite{ongoing} show that the predicted $|x''|$ envelope matches the true value with under 25\% per-sample relative error, supporting its use for adaptive knot placement.

\begin{figure}[p]
    \centering
    \includegraphics[width=0.98\linewidth]
    {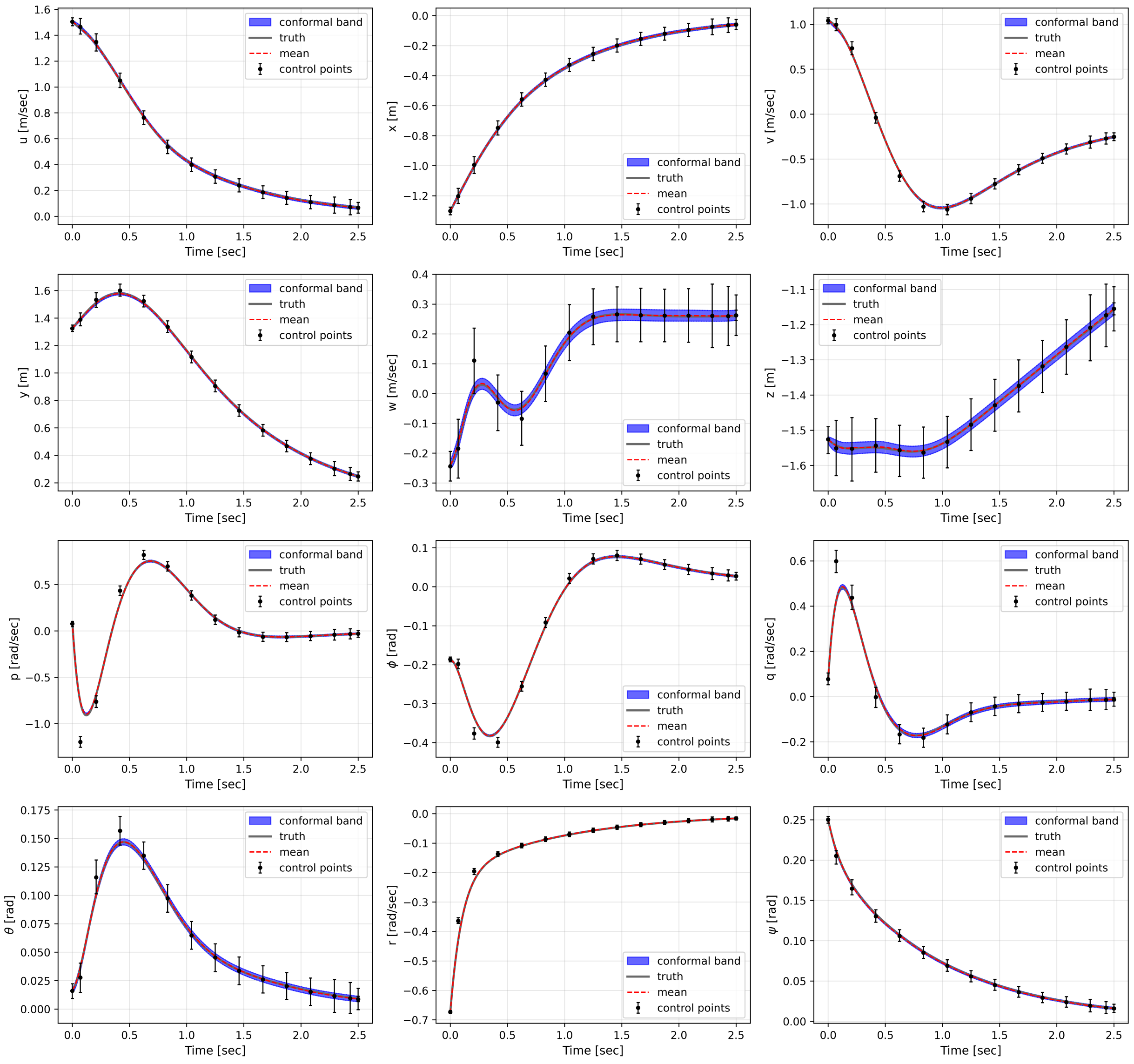}
    \caption{Same as Figure~\ref{fig:control_points_full} with the same
    number of knots, but now with a uniform distribution. Notably, the
    control intervals are wider and farther from the trajectory.}
    \label{fig:even_knots}
\end{figure}

Figure~\ref{fig:control_points_full} shows the control points found for the given trajectory using this method. Figure~\ref{fig:even_knots} shows the same result and trajectory but with the same number of knots uniformly spaced. Here one can see that the evenly spaced knots return control point intervals that are wider and farther from the trajectory.

\clearpage

\section{Recurrent Neural Network Comparison}
\label{app:rnn}
\begin{figure}
    \centering
    \includegraphics[width=0.95\linewidth]{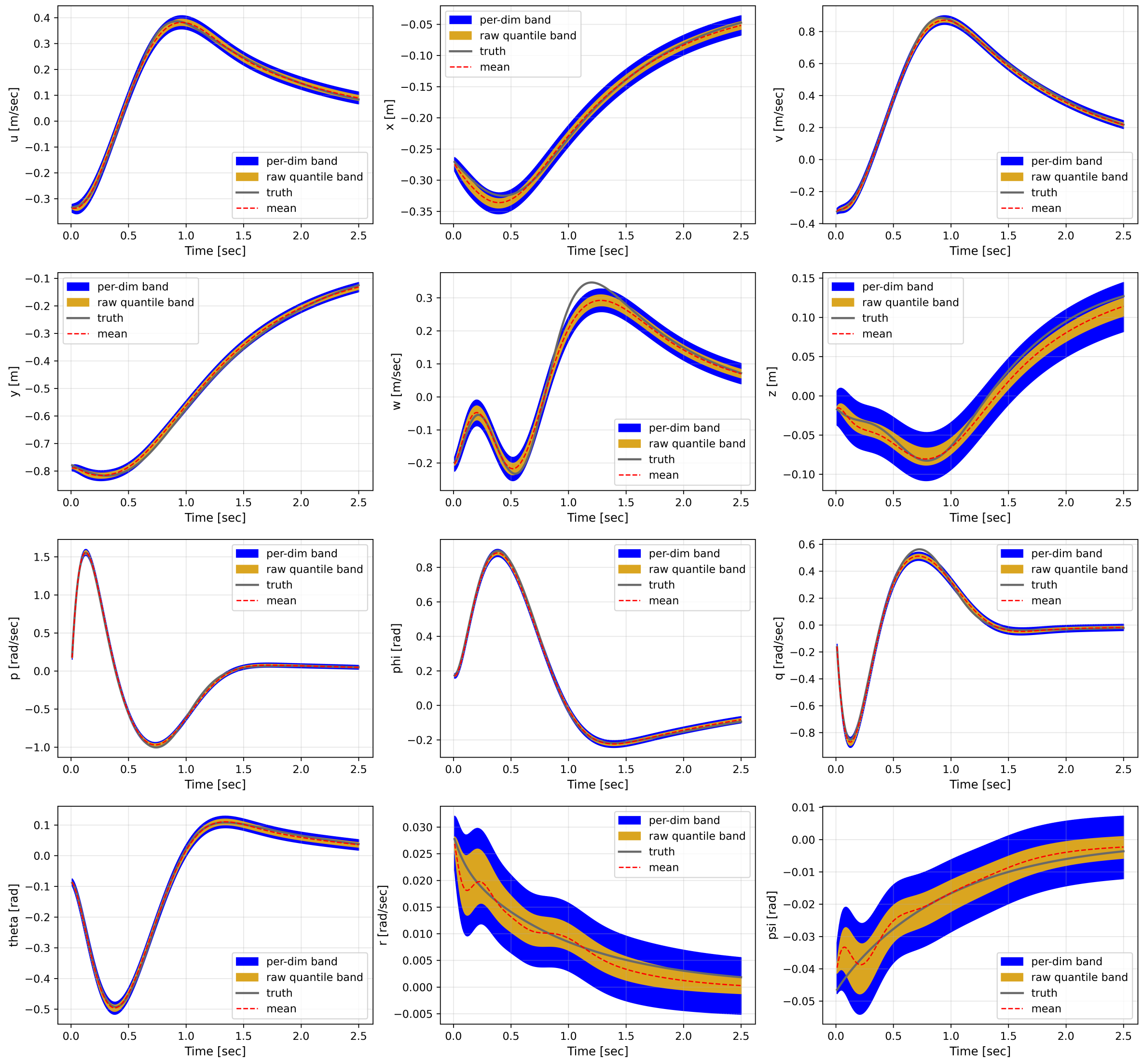}
    \caption{Using an RNN instead of the spectral backbone. Compare to Figure~\ref{fig:cqr95_full} and Figure~\ref{fig:cqr99_full}. Similar to those figures, this represents the ``worst'' hold-out sample based on its accuracy prior to CQR.}
    \label{fig:rnn}
\end{figure}

We replace the spectral backbone with a size-matched recurrent neural network (RNN)  (33,232 vs. 33,024 parameters), sharing the same lift/decode interface. In order to match both interfaces and parameter size, the \emph{hidden layer} has a size of 72. The input size (128) and output size (256) are determined by the interfaces. The RNN used has no long-term memory. This is unnecessary here because the system is Markovian with no long-range dependency structure. We address gradient stability over the long unroll (T = 251) by using a near-identity, slightly contractive cell initialization with Tanh, rather than gating, and by reducing the learning rate from 0.001 to 0.0001.  The RNN version is shown in Figure~\ref{fig:rnn}. The spectral backbone version converges to a lower loss and has tighter CQR bands.
In addition, we note that a recurrent architecture has several deficiencies unrelated to accuracy. For example, an RNN is slower at inference and cannot avail itself of the "tricks" we used to accelerate the inference on the Raspberry Pi. Furthermore, the RNN speed is sensitive to \emph{T} (the number of rolled out time steps) and could exhibit compounding errors at long time horizons. We considered a feed-forward implementation, but just the output layer --- with $\text{lifted dimension } \times \text{Rolled Out Time Steps }  = 128 \times 251 = 32128 $ --- would require nearly all of the learnable parameters. From Table~\ref{tab:cqr_rnn}, we see that the mean post-CQR width is 0.035, or about 4.7 times larger than the post-CQR width using the spectral backbone.

\begin{table}[hbtp]
\centering
\caption{Conformal quantile regression on the RNN model ($\ell{=}128$, $m{=}251$, seed 42) trained for the 95\% coverage target.}
\label{tab:cqr_rnn}
\vspace{0.2in}
\setlength{\tabcolsep}{5pt}
\begin{tabular}{lccccccc}
\hline
\textbf{Target} & \multicolumn{2}{c}{\textbf{Raw coverage}} & \multicolumn{2}{c}{\textbf{Post-CQR coverage}} & \textbf{Raw width} & \textbf{Post-CQR width} & \textbf{$Q_\eta$} \\
 & cell & joint & cell & joint & (mean) & (mean) & \\
\hline
95\% & 87.61\% & 0.00\%  & 99.95\% & 94.77\% & 0.01076 & 0.03518 & 0.00537 \\
\hline
\end{tabular}

\vspace{0.5em}

\begin{tabular}{lcccccccccccc}
\hline
\textbf{Target} & $u$ & $x$ & $v$ & $y$ & $w$ & $z$ & $p$ & $\phi$ & $q$ & $\theta$ & $r$ & $\psi$ \\
\hline
95\% & 0.827 & 0.349 & 0.905 & 0.323 & 1.106 & 0.739 & 1.975 & 2.959 & 1.820 & 2.265 & 0.459 & 1.496 \\
\hline
\end{tabular}
\end{table}

\clearpage
\section{Initial Condition Uncertainty Propagation}
\label{app:ICUP}

Consider an initial condition $x_0$ with uncertainty, assumed to be Gaussian ($N(\mu=x_0,\sigma_0)$). Sampling and computing trajectories for a ball of initial condition values near $x_0$ is embarrassingly parallel, and the combined set of quantiles can be used to create a marginal probability and uncertainty on the upper and lower quantiles of the trajectory. The perturbation set is a single batched forward pass; we assume knots are fixed based on $x_0$. For a given bound $y$ and sampled points from the distribution $x'$, we use

\[
p(y \mid x_0) = \int p(y \mid x')\,p(x' \mid x_0)\,dx'
\]

to develop probabilistic upper and lower quantiles. This approach combines uncertainty from measurement noise with model uncertainty to create a reasonable guess of how initial condition uncertainty affects the trajectory's upper and lower quantile bands. An illustrative example is shown below in Figure~\ref{fig:robustness_u}. Note that the magnitude of the noise in each dimension was set arbitrarily (near 0.01), and the true effect of initial condition uncertainty depends on the assumed covariate shift. We set aside a full covariate-shift analysis based on \citet{tibshirani2019conformal} as future work.

\begin{figure}
    \centering
    \includegraphics[width=0.98\linewidth]{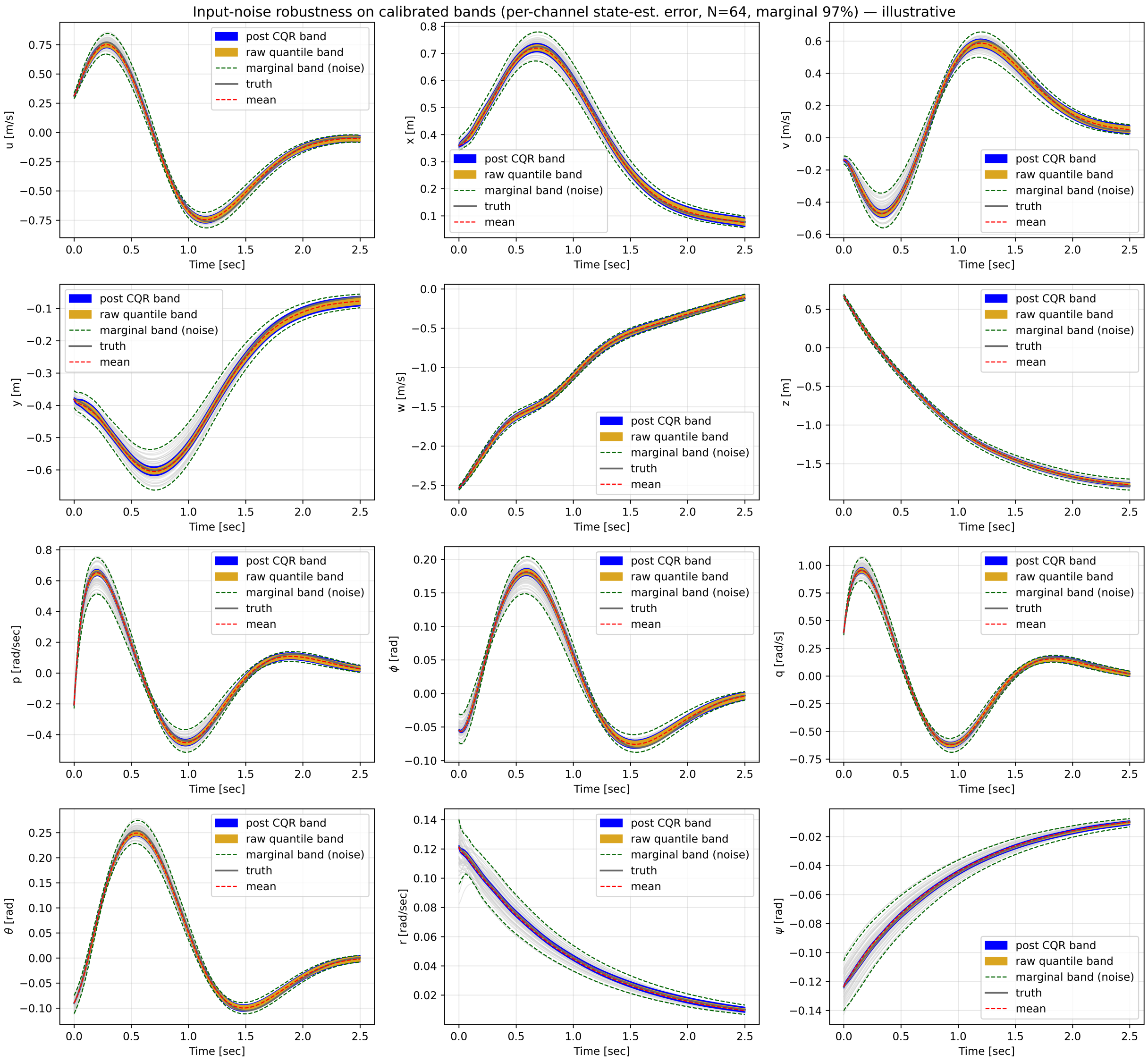}
    \caption{Input-noise propagation for one initial condition with a small amount of noise in each dimension. Black is truth; red dashed is the mean prediction; gold is the raw quantile band; blue is the per-dim conformal-calibrated band; green dashed is the marginal band from perturbing the IC with representative sensor-noise magnitudes and taking mixture quantiles across the perturbation set. This does \emph{not} provide coverage guarantees in the sense of \citet{tibshirani2019conformal}, and we intend a full study with a justified model for the distribution shift.}
\label{fig:robustness_u}
\end{figure}

\end{document}